\documentclass[aps, prb, twocolumn, twoside, 10pt, letterpaper,
superscriptaddress]{revtex4}

\usepackage{graphicx}
\usepackage{bbm,bm}
\usepackage{amssymb, amsmath,mathtools}
\usepackage{slashed}
\usepackage[colorlinks=true, urlcolor=blue, linkcolor=black,%
  citecolor=blue]{hyperref}
\newcommand{\Nf}{{N_\mathrm{f}}}                                      
\DeclareMathOperator{\Tr}{Tr}                                                 
\DeclareMathOperator{\STr}{STr}				              
\begin{document}
\graphicspath{{./}{../plots/}}
\title{Antiferromagnetic critical point on
graphene's honeycomb lattice:\\ A functional renormalization group approach}
\author{Lukas Janssen}
\email{lukasj@sfu.ca}
\affiliation{Department of Physics, Simon Fraser University,
Burnaby, British Columbia, Canada V5A 1S6}
\affiliation{Theoretisch-Physikalisches Institut,
Friedrich-Schiller-Universit{\"a}t Jena, Max-Wien-Platz 1, 07743 Jena, Germany}
\author{Igor F.\ Herbut}
\affiliation{Department of Physics, Simon Fraser University,
Burnaby, British Columbia, Canada V5A 1S6}
\affiliation{Max-Planck-Institut f\"ur Physik komplexer Systeme, N\"othnitzer
Str.\ 38, 01187 Dresden, Germany}
\begin{abstract}
Electrons on the half-filled honeycomb lattice are expected to undergo a direct continuous transition from the semimetallic into the
antiferromagnetic insulating phase with increase of on-site Hubbard repulsion. We attempt to further
quantify the critical behavior at this quantum phase
transition by means of functional renormalization group (RG), within an
effective Gross-Neveu-Yukawa theory for an $\mathrm{SO}(3)$ order parameter
(``chiral Heisenberg universality class''). Our calculation yields an estimate of the critical exponents $\nu
\simeq 1.31$, $\eta_\phi \simeq 1.01$, and $\eta_\Psi \simeq 0.08$, in reasonable agreement with the second-order
expansion around the upper critical dimension. To test
 the validity of the present method we use the conventional
Gross-Neveu-Yukawa theory with $\mathbbm Z_2$ order parameter (``chiral Ising
universality class'') as a benchmark system. We explicitly show that our
functional RG approximation in the sharp-cutoff scheme becomes one-loop exact
both near the upper as well as the lower critical dimension. Directly in $2+1$
dimensions, our chiral-Ising results agree with the best available predictions
from other methods within the single-digit percent range for $\nu$ and
$\eta_\phi$ and the double-digit percent range for $\eta_\Psi$. While one would expect
a similar performance of our approximation in the chiral Heisenberg universality
class, discrepancies with the results of other calculations here are more significant.
Discussion and summary of various  approaches is presented.
\end{abstract}

\maketitle

\section{Introduction}
Graphene is an excellent conductor. Experiments show that this remains true even
for suspended graphene sheets, when the substrate is removed.\cite{2011NatPh...7..701E}
Recent accurate \emph{ab initio} computations of the
strength of Coulomb repulsion in free-standing graphene, however, find values
which would place graphene not too far from the quantum phase transition into a putative
Mott-insulating phase.\cite{2011PhRvL.106w6805W, PhysRevLett.111.056801}
It is thus not inconceivable that there exist situations in which the Coulomb
interaction between the electrons would become strong enough relative to the bandwidth, so that a band gap in
the electronic spectrum is dynamically generated.
Such an effect may, for example, be observed in mechanically stretched graphene
sheets, where the hopping of the electrons between neighboring sites would (albeit most likely non-uniformly) be reduced.
Tuning through a semimetal--Mott-insulator phase transition could facilitate extraordinary
applications for graphene-based electronics, and would therefore be also highly
desirable from a technological point of view. On the other hand, because of its
Dirac-type spectrum, a Mott transition in graphene mimics the spontaneous
symmetry breakdown in high-energy particle physics, as it occurs in the strong
and electroweak sectors. Understanding the correlated physics of graphene near
criticality can therefore, as it has already, fertilize further the research on
some of the most intriguing issues of modern fundamental
physics: chiral symmetry breaking in QCD, the electroweak phase transition and
the Higgs mechanism, and the triviality problem in asymptotically nonfree
sectors of the standard model of particle physics.

The nature of the quantum phase transition on gra\-phene's honeycomb lattice has been under much
debate.\cite{tossati,PhysRevLett.97.146401, meng} Recently, however, the results began to converge towards the scenario with a
single second-order phase transition between the semimetallic and Mott-insulating states: Analytical results for all perturbatively accessible deformations of the theory
near $1+1$ \cite{PhysRevB.80.081405} and $3+1$ \cite{Herbut:2009vu} dimensions,
and in the $1/N$ \cite{PhysRevLett.97.146401, PhysRevB.75.235423} expansion,
suggest that the strength of the long-range part ($\sim 1/r$) of the Coulomb interaction, at least
when not too strong and at accessible length scales, is a marginally
irrelevant coupling, and that the transition is triggered by strong short-range components of the
interaction. \cite{PhysRevB.80.081405}
For the Hubbard model on the honeycomb lattice, recent quantum Monte Carlo (MC) calculations find for strong on-site repulsion a
direct and continuous quantum phase transition into the antiferromagnetic
insulator.\cite{2012NatSR...2E.992S, Assaad:2013xua}
Universality suggests that
the transition should be within the $\mathrm{SU}(2)$-Gross-Neveu (``chiral
Heisenberg'') universality class and the scaling behavior of the MC data indeed fits
persuasively well to the predictions from the first-order $\epsilon$-expansion of the
$\mathrm{SU}(2)$-Gross-Neveu-Yukawa field theory.\cite{Assaad:2013xua}
A reliable calculation
of the critical exponents is, as always, a challenging task and---very similar to
the much investigated bosonic $\mathrm O(N)$ universality classes---accurate
numerical estimates for the universal quantities in $2+1$ dimensions can only be
obtained by convergence of results from several complementary approaches.\cite{Herbut2007modern}
However, besides the $\epsilon$-expansion results
\cite{Rosenstein:1993zf, Herbut:2009vu} and the quantum Monte Carlo on honeycomb lattice,\cite{2012NatSR...2E.992S, Assaad:2013xua}
there are to date no other predictions for the critical exponents of the chiral Heisenberg
universality class available.

The aim of this article is therefore to attempt to further quantify the critical behavior of the
chiral Heisenberg universality class by means of functional renormalization group (RG) methods.
The functional RG has successfully been used to describe a variety of different
correlated fermion systems.\cite{RevModPhys.84.299, Platt:2013ada} In the context of graphene
it has been employed to determine the dominant instabilities on
single-layer,\cite{PhysRevLett.100.146404} bilayer,\cite{PhysRevB.85.235408} and
trilayer\cite{PhysRevB.86.155415} honeycomb lattices
at and away \cite{Kiesel:2012} from half filling. By taking collective
(Hubbard-Stratonovich--type) degrees of freedom into account, the functional RG
has been shown to be an excellent tool to   describe $(2+1)$-dimensional relativistic fermion
models at criticality.\cite{PhysRevLett.86.958, Hofling:2002hj,
PhysRevD.83.085012, Braun:2011pp, PhysRevD.81.025009, PhysRevD.82.085018,
PhysRevD.86.105007, PhysRevB.86.245431}

We first use the $\mathbbm Z_2$-Gross-Neveu (``chiral Ising'') universality class as a
benchmark system to estimate the validity of our approximation. The chiral Ising
universality class is supposed to describe the transition into a ``charge density wave" (CDW)
phase, with a broken sublattice symmetry, favored by a large nearest-neighbor repulsion on the honeycomb
lattice.\cite{PhysRevLett.97.146401, PhysRevLett.100.146404}
Critical exponents have been computed to 3rd loop order near $D=1+1$ space-time
dimensions (with the anomalous dimensions up to 4th order),\cite{Gracey:1990sx,
Gracey:1991vy, Luperini:1991sv, Vasiliev:1997sk, Gracey:2008mf} to 2nd order
near $D=3+1$ dimensions,\cite{Karkkainen:1993ef, Rosenstein:1993zf} to 2nd
order in the $1/\Nf$-expansion (with the fermionic anomalous dimension up to 3rd
order),\cite{Vasiliev:1992wr, Gracey:1993kc} using Monte-Carlo simulations,
\cite{Karkkainen:1993ef} as well as functional RG methods.
\cite{PhysRevLett.86.958, Hofling:2002hj, PhysRevD.83.085012}

The loop corrections in the expansions are generically only slowly (or even not
at all) decreasing with the order, such that the naive extrapolation to the
physical case with $\epsilon = 1$ and/or $\Nf = 2$ is often not
without problems. Due to the lack of knowledge of the large-order behavior of
the coefficients, standard Borel-type resummation techniques appear to be
hardly justified. We argue that a sensible resummation of the $\epsilon$-expansions
can be obtained by using the information from the $(2+\epsilon)$-expansion and
the $(4-\epsilon)$-expansion \emph{simultaneously} in terms of an interpolation
between those two limits. For the chiral Ising universality class, we show that
our functional RG results in the sharp-cutoff scheme become one-loop exact both
near the upper as well as the lower critical dimension, and that for a general dimension
$2<D<4$ they agree remarkably well with the proposed interpolational resummations. They also
agree with the predictions from all other methods within the mid single-digit
percent range for $\nu$ and $\eta_\phi$ and the lower double-digit percent range
for $\eta_\Psi$; see Table~\ref{tab:exponents_chiral_Ising}.

For the chiral Heisenberg universality class, which is assumed to describe the
antiferromagnetic phase transition on the honeycomb lattice, much fewer results
are available at the moment. With\-in the functional RG approach we obtain estimates for the
critical behavior in terms of the correlation length exponent $\nu$, the
anomalous dimensions for order parameter $\eta_\phi$ and for the fermionic field
$\eta_\Psi$, as well as the corrections-to-scaling exponent $\omega$; see
Table~\ref{tab:exponents_chiral_Heisenberg}. Our results for $\nu$ and $\eta_\phi$
agree reasonably well with the  previous second-order $\epsilon$-expansion, whereas the
result for  $\eta_\Psi$ is significantly different. The results in $D=2+1$ are
also numerically quite different from the
lowest-order $\epsilon$-expansion,\cite{Herbut:2009vu} which on the other hand, agrees surprisingly well with the Monte Carlo study of the
Hubbard model on honeycomb lattice.\cite{Assaad:2013xua} More numerical and analytical studies of this universality class are obviously needed.

The rest of the article is organized as follows:
In the next section we describe our effective model, its symmetries and breaking
patterns. A brief introduction to the functional RG approach is given in
Sec.~\ref{sec:functional_RG}, and the flow equations are derived in
Sec.~\ref{sec:flow_equations}. In Sec.~\ref{sec:fixed_points} we discuss the
fixed points first by expanding around the upper critical dimension, thereby
confirming previous Wilsonian RG $\epsilon$-expansion results, and eventually by
numerically evaluating the full set of flow equations for general space-time
dimension $2<D<4$. We discuss our results and compare them extensively with the
existing literature in Sec.~\ref{sec:discussion}. Conclusions are presented in
Sec.~\ref{sec:conclusions}.

\section{Effective theory}
The spin-$1/2$ electrons on the honeycomb lattice are described by the
$8$-component Dirac fermion fields
\mbox{$\Psi =
\left(\begin{matrix}
\Psi_\uparrow \\ \Psi_\downarrow
\end{matrix}\right)$} and its Dirac conjugate $\bar\Psi = \Psi^\dagger
(\mathbbm{1}_2 \otimes \gamma_0)$
in $2<D<4$ space-time dimensions. $\Psi_{\uparrow,\downarrow}$ denote the two
four-component spinors for direction up and down of the physical spin.
Due to the increase of the Fermi velocity $v_\mathrm F$ near half filling,\cite{2011NatPh...7..701E} the weak long-range part of the static Coulomb interaction (effective graphene fine-structure constant) appears to be an irrelevant coupling, and near the criticality Lorentz
invariance is emergent. \cite{PhysRevB.80.081405, Herbut:2009vu,
PhysRevLett.97.146401, PhysRevB.75.235423} The divergence of $v_\mathrm F$, of course, is an artifact of the static model, and the Fermi velocity ultimately can not exceed the velocity of light. The critical exponents we compute in the following will thus in principle receive corrections of the order of the QED fine-structure constant $\simeq 1/137$. \cite{Gonzalez:1993uz}
In any realistic experimental or numerical setup the running of $v_\mathrm F$ is bound by finite temperature and the  system's size.
In our model we will henceforth ignore these corrections, and retain only the short-range parts of the Coulomb repulsion. \cite{PhysRevB.80.081405}
The Euclidean effective theory describing the Mott transition
with integrated-out Coulomb field
is then explicitly relativistic; it is given in terms of $\Psi$,
$\bar\Psi$, and the order parameter field~$\phi_a$ as\cite{Herbut:2009vu}
\begin{align} \label{eq:action}
\mathcal S = \int \mathrm{d}\tau \mathrm{d}^{D-1} \vec{x} \biggl[ \bar \Psi
(\mathbbm{1}_2 \otimes \gamma_\mu) \partial_\mu \Psi
+ \frac{1}{2} \phi_a \left( \bar m^2 - \partial_\mu^2 \right) \phi_a \nonumber
\\
+ \bar\lambda \left(\phi_a^2\right)^2 + \bar g \phi_a \bar\Psi (\sigma_a \otimes
\mathbbm{1}_4) \Psi\biggr],
\end{align}
with the space-time index $\mu=0,1,\dots,D-1$, the $D$-derivative
$(\partial_\mu) = (\partial_\tau,\vec\nabla)$ and the $4\times 4$ gamma
matrices, obeying the Clifford algebra $\{\gamma_\mu,\gamma_\nu\} =
2\delta_{\mu\nu}$. Summation over repeated indices is assumed. The
overbar emphasizes the dimensionfulness of the coupling constants $\bar\lambda$
and $\bar g$. In the direct products $\sigma_a \otimes \gamma_\mu$ the Pauli
matrices act on spin, and the gamma matrices act on Dirac  indices. The index $a$
either runs from $1$ to $3$, to which we will refer to as ``chiral Heisenberg''
\cite{Rosenstein:1993zf} model in the following, or it is fixed $a\equiv 0$ with
$\sigma_0 \equiv \mathbbm{1}_2$. We will refer to the latter case as
``chiral Ising'' \cite{Rosenstein:1993zf} model.
The standard Ising and Heisenberg universality classes can be recovered from the chiral models by artificially setting $\bar g \equiv 0$. Our chiral systems thus agree with their purely bosonic (non-chiral) counterparts in terms of the order-parameter symmetry. They differ, however, in that they incorporate massless (chiral) fermionic modes, and they thus describe different universality classes.
In $2+1$ dimensions we may use the ``graphene'' representation
\cite{PhysRevLett.97.146401} $\gamma_0 = \mathbbm{1}_2 \otimes \sigma_z$,
$\gamma_1 = \sigma_z \otimes \sigma_y$, and $\gamma_2 = \mathbbm{1}_2 \otimes
\sigma_x$. In this representation the Grassmann fields $u$ and $v$ on the two
sublattices of the honeycomb lattice near the Dirac point $\vec K$ are related
to the Dirac field as
\begin{align}
\Psi_\sigma^\dagger(\vec x,\tau) & = \int
\frac{\mathrm{d}\omega \,\mathrm{d}^{D-1}\vec q}{(2\pi)^D} e^{i \omega
\tau + i \vec q \cdot \vec x} \Bigl[ u^\dagger_\sigma(\vec K + \vec q, \omega),
\nonumber \\ & \!
v^\dagger_\sigma(\vec K + \vec q,\omega), u^\dagger_\sigma(-\vec K + \vec
q,\omega), v^\dagger_\sigma(-\vec K + \vec q,\omega) \Bigr],
\end{align}
where we have chosen a reference frame in which $q_x = \vec q \cdot \vec K / |K|$,
 and for simplicity have set the lattice spacing and the Fermi velocity to unity.
There are two further $4\times 4$ matrices which anticommute with all three
$\gamma_\mu$: $\gamma_3 = \sigma_x\otimes\sigma_y$ and $\gamma_5 =
\sigma_y\otimes \sigma_y$. The Hermitian product $\gamma_{35} =
-i\gamma_3\gamma_5$ commutes with the $\gamma_\mu$'s and anticommutes with
$\gamma_3$ and $\gamma_5$. Note that it is diagonal in our representation.

Let us discuss the symmetries of our effective relativistic models and relate
them to the structure of the underlying honeycomb lattice. The action in
Eq.~\eqref{eq:action} exhibits a discrete reflection symmetry,
\begin{align}
\mathbbm{Z}_2: \
\Psi \mapsto (\mathbbm 1_2 \otimes \gamma_2) \Psi, \
\bar\Psi \mapsto -\bar\Psi (\mathbbm 1_2 \otimes \gamma_2), \
\phi_a \mapsto -\phi_a,
\end{align}
with the (spatial) momentum reflected across the first axis: $q_x \mapsto q_x$,
$q_y \mapsto - q_y$. Again, $a\equiv 0$ in the chiral Ising model and $a=1,2,3$
in the chiral Heisenberg model, respectively. This defines the
sublattice-exchange symmetry of the honeycomb lattice, which exchanges the two
Grassmann fields $u \leftrightarrow v$.\cite{Herbut:2009qb}
Both models are furthermore invariant under
$\mathrm{SU}(2)$ spin rotations, under which $\phi_0$ is a scalar and $\vec \phi
= (\phi_a)_{a=1,2,3}$ transforms as a vector:
\begin{align}
\mathrm{SU}(2)_\text{sp}&: & \Psi &\mapsto e^{i\theta \vec n \cdot
(\vec\sigma \otimes \mathbbm{1}_4)}\Psi, & \bar\Psi & \mapsto \bar\Psi e^{-i
\theta \vec n \cdot (\vec\sigma \otimes \mathbbm{1}_4)}, \nonumber\\ &
 & \phi_0 & \mapsto \phi_0, & \vec \phi & \mapsto R \vec\phi, &
\end{align}
with rotation matrix
$(R_{ab}) = (\delta_{ab} - 2\theta\epsilon_{abc} n_c ) \in
\mathrm O(3)$.
Here, we have used
$[(\sigma_a \otimes \mathbbm 1_4), (\sigma_b \otimes \mathbbm 1_4)] = 2 \epsilon_{abc} (\sigma_c \otimes \mathbbm 1_4)$, ensuring that the chiral Heisenberg bilinear $\bar\Psi (\vec \sigma \otimes \mathbbm 1_4)\Psi$ transforms as a vector under $\mathrm{SU}(2)_\text{sp}$.
Charge conservation requires the usual
$\mathrm U(1)_\text{ch}$ phase-rotational symmetry $\Psi \mapsto e^{i \theta} \Psi$,
$\bar\Psi \mapsto \bar\Psi e^{-i \theta}$. However, the charge in each
Dirac-cone sector at wavevectors $\pm \vec K$ is conserved separately, and the
phases of the modes in the two valleys can therefore be rotated independently.
Formally, this can be seen by making use of the ``chiral'' projector $P_\pm =
\mathbbm{1}_2 \otimes (\mathbbm{1}_4 \pm \gamma_{35})/2$, which projects onto
the modes near $\pm \vec K$. The corresponding ``chiral'' $\mathrm U(1)$ symmetry
is
\begin{align}
\mathrm{U(1)}_\chi &: &
\Psi & \mapsto e^{i \theta (\mathbbm{1}_2 \otimes \gamma_{35})} \Psi, &
\bar\Psi & \mapsto \bar\Psi e^{-i \theta (\mathbbm{1}_2 \otimes \gamma_{35})}.
\end{align}
On the honeycomb lattice, $\mathrm U(1)_\chi$ in fact corresponds to translational
invariance.\cite{Herbut:2009qb}
Additional to the phase rotations, in the chiral Ising model the two modes at
$\pm \vec K$ can also be rotated independently in spin space. The chiral
symmetry here is thus elevated to $ \mathrm U(2)_\chi \simeq \mathrm{U}(1)_\chi \times
\mathrm{SU}(2)_\chi$, with
\begin{align} \label{eq:chiral_ising_2nd-SU2}
 \mathrm{SU}(2)_\chi&: &
\Psi &\mapsto e^{i\theta \vec n \cdot (\vec\sigma \otimes \gamma_{35})}\Psi, &
\bar\Psi & \mapsto \bar\Psi e^{-i \theta \vec n \cdot (\vec\sigma \otimes
\gamma_{35})},
\end{align}
while keeping the order-parameter field $\phi_0 \mapsto \phi_0$ fixed.
In the chiral Heisenberg model, however, since the commutator $[(\sigma_a \otimes \mathbbm \gamma_{45}), (\sigma_b \otimes \mathbbm 1_4)]$ is \emph{not} proportional to $\sigma_c \otimes \mathbbm 1_4$, the bilinear $\bar\Psi (\vec \sigma \otimes
\mathbbm{1}_4) \Psi$ is not a vector under $\mathrm{SU}(2)_\chi$. Hence, the chiral symmetry here is not elevated, and remains $\mathrm U(1)_\chi$.
Altogether, the symmetry groups of the
chiral Ising and the chiral Heisenberg model therefore are
\begin{align}
&\text{$\chi$-Ising}: && \mathbbm{Z}_2 \times \mathrm{SU}(2)_\text{sp} \times
\mathrm{U}(1)_\text{ch} \times \mathrm{U}(2)_\chi, \\
&\text{$\chi$-Heisenberg}: && \mathbbm{Z}_2 \times \mathrm{SU(2)}_\text{sp}
\times \mathrm{U}(1)_\text{ch} \times \mathrm{U}(1)_\chi.
\end{align}

For strong coupling the order-parameter field can develop a nonvanishing vacuum
expectation value (VEV). In the chiral Ising case with a single order-parameter
field ($a\equiv 0$) a VEV $\langle \phi_0 \rangle \propto \langle \bar\Psi\Psi
\rangle \neq 0$ breaks the $\mathbbm Z_2$ sublattice-exchange symmetry
spontaneously, and our model describes the second-order transition into the
staggered-density phase, the charge density wave (CDW) state. The critical
behavior is described by the celebrated $\mathbbm{Z}_2$-Gross-Neveu ($=$ chiral
Ising) universality class, the corresponding universal exponents being fairly
well known.\cite{Gracey:1990sx, Gracey:1991vy, Luperini:1991sv, Vasiliev:1997sk,
Gracey:2008mf, Vasiliev:1992wr, Gracey:1993kc, Rosenstein:1993zf,
Karkkainen:1993ef, Herbut:2009vu, PhysRevLett.86.958, Hofling:2002hj,
PhysRevD.83.085012, Braun:2011pp}
In contrast, the chiral Heisenberg model with the $3$-vector order-parameter
field $\vec \phi = (\phi_1,\phi_2,\phi_3)$ describes the transition of the
semimetallic phase into the staggered-magnetization state, the antiferromagnetic
(AFM) phase. If $\vec\phi$ develops a VEV, $\langle \vec \phi \rangle \propto
\langle \bar\Psi (\vec \sigma \otimes \mathbbm{1}_4) \Psi \rangle \neq \vec 0$,
both the $\mathbbm Z_2$ sublattice-exchange symmetry as well as the
$\mathrm{SU(2)}_\text{sp}$ spin-rotational symmetry are spontaneously broken
down to a residual $\mathrm{O}(2) \simeq \mathrm{U}(1)$ symmetry. On the AFM
side of the transition
we therefore expect $2$ massless bosonic modes, the Goldstone modes,
corresponding to the field variables being orthogonal to the VEV. The
corresponding chiral Heisenberg [$=$ $\mathrm{SU}(2)$-Gross-Neveu] universality
class is not so well understood (see, however, Refs.~\onlinecite{Rosenstein:1993zf,
Herbut:2009vu} for results within an expansion around the upper critical
dimension). In the following, we will investigate both the chiral Ising and the
chiral Heisenberg universality classes by means of the functional renormalization
group.

\section{Functional renormalization group} \label{sec:functional_RG}
The functional renormalization group (FRG) approach is an efficient tool to
compute the generating functional of the one-particle irreducible correlation
functions---the effective action $\Gamma[\phi_a,\Psi,\bar\Psi]$%
\footnote{Note that for notational simplicity we use the same symbols for the
fluctuating fields $\Psi$, $\bar\Psi$, $\phi_a$ and the arguments of $\Gamma$,
i.e., the field expectation values ${\langle}\phi_a
{\rangle}_{j,\eta,\bar\eta}$, ${\langle}\Psi {\rangle}_{j,\eta,\bar\eta}$, and
${\langle}\bar\Psi {\rangle}_{j,\eta,\bar\eta}$ in the presence of the
conjugated sources $j$, $\eta$, and $\bar\eta$.}.
For reviews on this rapidly evolving method, applied to both condensed-matter as
well as high-energy physics, see Refs.~\onlinecite{kopietz2010introduction, Berges:2000ew,
Aoki:2000wm, Polonyi:2001se, Pawlowski:2005xe, Gies:2006wv, Delamotte:2007pf,
Sonoda:2007av, RevModPhys.84.299, Platt:2013ada, Braun:2011pp}.
A thorough and very pedagogical introduction can be found in Ref.~\onlinecite{Wipf:2013vp}.
The central object of the method is the scale-dependent \emph{effective average
action} $\Gamma_k[\phi_a,\Psi,\bar\Psi]$, which is essentially the Legendre
transform of a regulator-modified action
\begin{align} \label{eq:action_regulator-modification}
\mathcal S \mapsto \mathcal S + \int \frac{\mathrm{d}^{D} q\mathrm{d}^D
p}{(2\pi)^{2D}} \biggl[ \frac{1}{2}\phi_a(-q) R^{(\mathrm B)}_{ab,k}(q,p)
\phi_b(p)
 \nonumber \\
 + \bar\Psi(q) R^{(\mathrm F)}_k (q,p) \Psi(p) \biggr],
\end{align}
with the bosonic regulator $R_k^{(\mathrm B)}(p,q) = \left(R_{ab,k}^{(\mathrm
B)}\right)(q,p)$, which for any given momenta $q$, $p$ is a $3\times 3$ matrix
in the chiral Heisenberg case ($a,b=1,2,3$) and a scalar in the chiral Ising
case ($a,b\equiv 0$), respectively; and the fermionic regulator $R_k^{(\mathrm
F)}(q,p)$, which is an $8 \times 8$ matrix acting on spin and Dirac indices.
Here, we have combined the frequency and momentum integration into the
integration over the relativistic $D$-momentum $q_\mu = (\omega,\vec q)$, with
space-time dimension $D$. In momentum space, the regulators, introduced here
integral kernels of linear operators in field space, are usually taken to be
diagonal, i.e., $R_k^\mathrm{(B/F)}(p,q) = R_k^\mathrm{(B/F)}(q) \delta(p-q)$.

At finite scale $k>0$, the regulator screens the IR fluctuations with $|q|\ll k$
in a mass-like fashion, ensuring that only fast modes with momentum $|q| \gtrsim
k$ give significant contributions to $\Gamma_k$. The fermionic regulator
$R_k^{(\mathrm F)}$ is constructed in a way that the regulator modification in
Eq.~\eqref{eq:action_regulator-modification} does not spoil the chiral symmetry.
Besides a sharp-cutoff regulator it is possible (and often very useful) to
employ smooth cutoff functions, which allow a continuous suppression of slow
modes.
For $k \to 0$ the regulator has to go to zero for all momenta, such that the
modifications in $\mathcal S$ vanish and the effective average action approaches
the full quantum effective action, $\Gamma_{k\to 0} = \Gamma$. We choose
regulator functions which for $k \to \Lambda$ are of the order of the UV cutoff
$\Lambda$, $R_{k\to\Lambda}^{(\mathrm B)} (q) \sim \Lambda^2$,
$R_{k\to\Lambda}^{(\mathrm F)} (q) \sim \Lambda$.
Thus, in the UV all fluctuations are suppressed and $\Gamma_{k\to\Lambda}$
becomes (up to normalization constants) the microscopic action,
$\Gamma_{k\to\Lambda} \simeq \mathcal S$. The effective average action thus
interpolates between the microscopic action in the UV and the full quantum
effective action in the IR.
The concept can be viewed as a specific implementation of Wilson's approach to
the renormalization group: Instead of integrating out all fluctuations at once,
we divide the functional integral into integrations over shells with momentum $q
\in [k,k - \delta k]$ and subsequently successively integrate momentum shell by
momentum shell. $\Gamma_k$ is the effective action at an intermediate step
$0\leq k \leq \Lambda$, where the fluctuations in the functional integral with
momentum $q \in [k,\Lambda]$ are integrated out. The theory then is solved, once
we know the evolution of $\Gamma_k$ with respect to the renormalization group
time $t = \ln (k/\Lambda)$ from $t=0$ (UV) to $t \to -\infty$ (IR). The
evolution equation for $\Gamma_k$ has been computed by Wetterich
\cite{Wetterich:1992yh} and is given by the functional identity
\begin{align} \label{eq:wetterich_eq}
\partial_t \Gamma_k = \frac{1}{2} \STr \left[\partial_t \mathrm R_k
\left(\Gamma_k^{(2)} + \mathrm R_k\right)^{-1} \right],
\end{align}
where $\mathrm R_k \coloneqq
\left(\begin{smallmatrix}
R_{k}^{(\mathrm{B})} & 0 &  0 \\
0 & 0 & R_k^{(\mathrm{F})} \\
0 & -R_k^{(\mathrm{F})T} & 0
\end{smallmatrix}\right)$
and $\Gamma_k^{(2)}$ denotes the second functional derivative of the effective
average action with respect to the fields $\phi_a$, $\Psi$, and $\bar\Psi$,
i.e.,
\begin{align}
\Gamma^{(2)}(p,q) \equiv \frac{\overrightarrow{\delta}}{\delta \Phi(-p)^T}
\Gamma_k \frac{\overleftarrow{\delta}}{\delta \Phi(q)},
\end{align}
where we have used the collective field variable $\Phi(q) =
\left(\begin{smallmatrix}
       \phi_a(q) \\ \Psi(q) \\ \bar\Psi(-q)^T
      \end{smallmatrix}
\right)$. Note that both $\mathrm R_k$ and $\Gamma_k^{(2)}$ define linear
operators acting on the collective field, e.g., $(\mathrm R_k \Phi)(p) \equiv
\int \frac{\mathrm d^D q}{(2\pi)^D} \mathrm R_k(p,q) \Phi(q)$.
$\STr$ runs over all internal degrees of freedom (momentum, spin, sublattice, valley),
as well as field degrees of freedom. In the fermionic sector, it takes an
additional minus sign into account,
$
\STr
     \left(\begin{smallmatrix}
       B & \ast & \ast \\
       \ast & F_1 & \ast \\
       \ast & \ast & F_2
      \end{smallmatrix}\right)
\coloneqq \Tr B -
\Tr
\left(\begin{smallmatrix}
       F_1 & \ast \\
       \ast & F_2
      \end{smallmatrix}\right)
$.

While the Wetterich equation \eqref{eq:wetterich_eq} is an exact identity for
the evolution of $\Gamma_k$, it is generically difficult to find exact
solutions. It is nevertheless perfectly possible to use it to find very
satisfying approximate solutions by means of suitable systematic expansion
schemes. Perturbation theory constitutes one such expansion; however, for the
description of phase transitions nonperturbative expansion schemes in terms of
operator or vertex expansions are often superior already at relatively low
order of the expansion. In particular, an expansion in terms of the derivative
has been shown to be highly suitable for the study of critical phenomena in
$(2+1)$-dimensional fermion-boson systems, yielding accurate predictions
for the critical exponents.\cite{PhysRevLett.86.958, Hofling:2002hj,
PhysRevD.81.025009, PhysRevD.83.085012, PhysRevD.86.105007, PhysRevB.86.245431}
In the spirit of the derivative expansion, we apply in this work the following
ansatz for the effective average action:
\begin{align} \label{eq:truncation}
\Gamma_k = \int \mathrm{d}^Dx \biggl[
Z_{\Psi,k} \bar\Psi \left( \mathbbm 1_{2} \otimes \gamma_\mu \right)
\partial_\mu \Psi
- \frac{1}{2} Z_{\phi,k} \phi_a \partial_\mu^2 \phi_a
\nonumber \\
 + U_k(\rho)
+\bar g_k \phi_a \bar \Psi \left( \sigma_a \otimes \mathbbm 1_4 \right) \Psi
\biggr],
\end{align}
with the scale-dependent wave-function
renormalizations $Z_{\phi,k}$, $Z_{\Psi,k}$ and the scale-dependent Yukawa-type
coupling $\bar g_k$.
For symmetry reasons, the scale-dependent effective bosonic potential $U_k$
has to be a function of the scalar product $\rho(x) \equiv \frac{1}{2}
\phi_a \phi_a$ only. It is often expanded in fields as
\begin{align} \label{eq:expansion_effective-potential}
U_k(\rho) = \sum_{n=1}^{\infty} \frac{\bar \lambda_k^{(n)}(0)}{n!} \rho^n,
\end{align}
with $\bar\lambda^{(1)}_k \equiv \bar m_k^2$ denoting the scalar-field mass.
This type of ansatz for $\Gamma_k$ is sometimes referred to as ``improved
local potential approximation'' (LPA').
The UV starting values for the flow are given by the microscopic couplings in
Eq.~\eqref{eq:action}, i.e.,
\begin{align}
 \lim_{k\to\Lambda} U_k(\rho) & = \bar m^2 \rho + 4 \bar\lambda \rho^2, &
 \lim_{k\to\Lambda} \bar g_k & = \bar g,
\end{align}
and
\begin{align}
 \lim_{k\to\Lambda} Z_{\phi,k} = \lim_{k\to\Lambda} Z_{\Psi,k} = 1.
\end{align}
At lower RG scales $k<\Lambda$, we absorb the wave-function renormalization factors $Z_{\phi/\Psi,k}$
into renormalized fields as
\begin{align}
 Z_{\phi,k}^{1/2} \phi_a & \mapsto \phi_a, &
 Z_{\Psi,k}^{1/2}\Psi & \mapsto \Psi, &
 Z_{\Psi,k}^{1/2}\bar\Psi & \mapsto \bar\Psi,
\end{align}
and use the dimensionless renormalized Yukawa-type coupling $g\equiv g(k)$
and dimensionless renormalized effective potential $u(\tilde \rho) \equiv
u(\tilde\rho;k)$:
\begin{align} \label{eq:dimensionless_couplings}
g^2 & = Z_{\phi,k}^{-1} Z_{\Psi,k}^{-2} k^{D-4} \bar g_{k}^2, &
u(\tilde \rho) & = k^{-D} U_k (Z_{\phi,k}^{-1} k^{D-2} \tilde\rho)
\end{align}
with $\tilde \rho = Z_{\phi,k} k^{2-D} \rho$. The anomalous dimensions
$\eta_{\phi/\Psi} = \eta_{\phi/\Psi}(k)$ are given by
\begin{align} \label{eq:anomalous_dimensions}
 \eta_\phi &  = - \frac{\partial_t Z_{\phi,k}}{Z_{\phi,k}} & \text{and} &&
 \eta_\Psi &  = - \frac{\partial_t Z_{\Psi,k}}{Z_{\Psi,k}}.
\end{align}

It should be worthwhile to discuss the approximations involved in our ansatz, Eq.~\eqref{eq:truncation}. In principle, all terms of higher order in derivative or fields being invariant under the present symmetry, could be generated under RG transformations. Schematically, they have the form
\begin{gather}
\label{eq:higher-order-ops_A}
{\bar \lambda}^{(m,n)}_{k} \partial^{2m} \phi^{2n},  \\
\label{eq:higher-order-ops_B}
{\bar h}^{(m,n)}_{k} \partial^{m}(\bar\Psi M \Psi)^n,  \quad
{\bar g_k}^{(m,n_1,n_2)} \partial^m \phi^{n_1} (\bar\Psi M \Psi)^{n_2},
\end{gather}
with suitable matrices $M \in \mathbbm C^{8\times 8}$.
In other words, even if we started the RG flow with pointlike coupling constants, the renormalized couplings could develop a momentum structure, i.e., we would have to deal with coupling \emph{functions} (in Fourier space); and, furthermore, new interactions could be generated, e.g., of the four-fermion type $(\bar\Psi M\Psi)^2$. The mass dimensions of these additional couplings are determined by
\begin{align}
\label{eq:scaling-dimension_higher-derivative_A}
[\bar \lambda_k^{(m,n)}] & = D - 2m - (D-2)n, \\
[\bar h^{(m,n)}_k] & = D - m -(D-1)n, \\
\label{eq:scaling-dimension_higher-derivative_C}
[\bar g^{(m,n_1,n_2)}_k] & = D - m - \frac{D-2}{2} n_1 - (D-1) n_2.
\end{align}
In $D>2$, all couplings neglected in our truncation of $\Gamma_k$ [Eq.~\eqref{eq:truncation}] thus have negative mass dimension. By contrast, the scaling dimensions of the couplings already present in our ansatz read as
\begin{align}
\label{eq:scaling-dimension_lambda-g}
[\bar \lambda^{(n)}_k] & = D - (D-2)n, &
[\bar g_k] & = \frac{1}{2} (4-D).
\end{align}
Below four space-time dimensions, $D<4$, $\bar\lambda^{(2)}$ and $\bar g_k$ thus have positive mass dimension, whereas they both become marginal directly in four dimensions. We thus recover \cite{Karkkainen:1993ef, Rosenstein:1993zf, Herbut:2009vu} that $D=4$ constitutes an upper critical dimension of the Gross-Neveu-Yukawa-type theories, and an anticipated critical point in $D=4-\epsilon$ would lie in the perturbatively accessible domain for small $\epsilon$. In this domain, however, the higher-derivative operators from Eqs.~\eqref{eq:higher-order-ops_A}--\eqref{eq:higher-order-ops_B} (as well as $\bar\lambda^{(n\geq 3)}_k$) are irrelevant in the RG sense, and we would be right to neglect them in our ansatz. Our truncation of $\Gamma_k$ will therefore become exact near $D=4$: To first order in $\epsilon$,  our
predictions for the critical exponents obtained by evaluating
Eq.~\eqref{eq:wetterich_eq} with the ansatz in Eq.~\eqref{eq:truncation} have to coincide exactly with
the known results from the $(4-\epsilon)$-expansion. \cite{Rosenstein:1993zf, Herbut:2009vu}
We will use this fact as a cross-check to verify the validity of our computation.

In the nonperturbative regime for not so small $\epsilon \sim \mathcal O(1)$, however, higher (perturbatively irrelevant) interactions
can be generated by the RG flow. Aside from higher-derivative terms,
higher bosonic self-interactions
$\propto (\phi_a^2)^n$, $n \geq 3$ may become important and might play a
quantitative role for the critical behavior. Below the UV cutoff scale, the
bosonic potential $U_k(\rho)$ therefore generically incorporates terms of
arbitrarily high order in $\rho$. This is an important advantage of the
functional RG approach: Contrary to conventional methods (e.g., within the
context of the $(4-\epsilon)$-expansion), it will prove possible to include all such
higher-order terms in $\rho$ by computing the full RG evolution of the effective
potential $U_k(\rho)$.
Moreover, in situations where different order parameters compete, the effect of newly generated four-fermion operators has been shown to play a decisive role. \cite{PhysRevD.82.085018} Within the present FRG scheme, they can be straightforwardly incorporated by the ``dynamical bosonization'' technique, \cite{Gies:2001nw} i.e., by performing a Hubbard-Stratonovich transformation at \emph{each} RG step. However, in our present partially bosonized models, with the single order parameter $\langle \phi_0 \rangle$ or $\langle \vec\phi \rangle$ at hand, we assume that four-fermion interactions do not become important at lower RG scales, and leave the dynamical bosonization technique for future improvement of our results.
In a next step, one can also go beyond LPA' by successively including the higher-derivative terms of Eqs.~\eqref{eq:higher-order-ops_A}--\eqref{eq:higher-order-ops_B} up to some fixed $m$. For the purely bosonic Ising universality class, various FRG studies have included terms up to 4th order in derivative, yielding predictions which agree with high-precision Monte-Carlo measurements, e.g., in terms of the correlation-length exponent $\nu$, within an error range of $\Delta \nu / \nu \simeq 0.5 \%$. \cite{Litim:2010tt}

\section{Flow equations} \label{sec:flow_equations}
\subsection{Bosonic potential}

The flow equation for the bosonic potential
$u(\tilde \rho)$ is readily obtained by plugging Eq.~\eqref{eq:truncation} into
the Wetterich equation~\eqref{eq:wetterich_eq}, and evaluating this functional
identity for constant bosonic field $\rho(x) = \rho = \text{const.}$, i.e.,
$\phi(p) = \phi \delta(p)$ in Fourier space, and vanishing fermionic field
$\Psi=\bar\Psi=0$. For this field configuration the regularized scale-dependent
two-point correlator $\Gamma^{(2)}_k + \mathrm R_k$ is block-diagonal and easily
inverted. We obtain for the chiral Ising (chiral Heisenberg) model with $S=0$
($S=2$) potential Goldstone modes:
\begin{align}
 \partial_t u(\tilde\rho) & =
-D u(\tilde \rho) + (D-2+\eta_\phi) \tilde\rho u'(\tilde \rho)
\nonumber \\ &\quad
+ 2 S v_D \ell_0^{\mathrm{(B)},D}\left(u'(\tilde \rho);\eta_\phi\right)
\nonumber \\ &\quad
+ 2 v_D \ell_0^{\mathrm{(B)},D}\left(u'(\tilde \rho)+2 \tilde\rho u''(\tilde
\rho);\eta_\phi\right)
\nonumber \\ &\quad \label{eq:dtu}
- 2 d_\gamma \Nf v_D \ell_0^{\mathrm{(F)},D}(2 \tilde\rho g^2; \eta_\Psi),
\end{align}
where we have introduced the dimensionless threshold functions
$\ell_0^{\mathrm{(B/F)},D}(\omega;\eta_{\phi/\Psi})$, which involve the
remaining loop integral and incorporate the dependence on the regulator function
$R_k^{\mathrm{(B/F)}}$. Their definitions are given in the Appendix.
$d_\gamma=\Tr(\gamma_0^2)$ is the size of the gamma matrices, and we have
abbreviated $v_D = (2\pi)^{-D} \text{vol}(S^{D-1})/4 =
1/\left(2^{D+1}\pi^{D/2}\Gamma(D/2)\right)$ with space-time dimension $D$. We
have also allowed for a general ``flavor'' number $\Nf$, counting the number of
electronic spin directions, with $\Nf=2$ in the physical case.

\subsection{Yukawa-type coupling}

In order to compute the beta function for the
Yukawa-type coupling $g$, we first rewrite Eq.~\eqref{eq:wetterich_eq} as
\begin{align}
\partial_t \Gamma_k = \frac{1}{2} \tilde \partial_t \STr \ln \left(
\Gamma_k^{(2)} + \mathrm R_k \right),
\end{align}
where the derivative $\tilde\partial_t$ is defined to act only on the
\emph{regulator's} $t$-dependence (and not on $\Gamma_k^{(2)}$), i.e.,
\begin{multline} \label{eq:tilde-partial_t}
\tilde \partial_t \coloneqq \int d^Dx' \left[
\partial_t R_k^\mathrm{(B)} (x') \frac{\delta}{\delta R_k^\mathrm{(B)} (x')}
\right. \\
\left.+\partial_t R_k^\mathrm{(F)} (x') \frac{\delta}{\delta R_k^\mathrm{(F)}
(x')}
\right] .
\end{multline}
Let $\kappa \equiv \tilde\rho_\text{min}$ be the value for which the effective
potential $u(\tilde\rho)$ at scale $k$ is at its minimum, $\partial_{\tilde\rho}
u|_{\tilde\rho=\kappa} = 0$. In the IR limit, $\kappa$ determines the field
expectation value $\langle \frac{1}{2}\phi_a \phi_a \rangle = \lim_{k \to 0}
k^{D-2} \kappa$.
Due to the fermionic fluctuations, which for our theory (with $2\Nf d_\gamma =
16$ fermionic degrees of freedom) will turn out to dominate the flow of the
effective potential, the RG fixed point corresponding to the anticipated
second-order phase transition is located in the symmetric regime, i.e., the
fixed-point potential $u^*(\tilde\rho)$ attains its minimum at the origin and
$\kappa$ exactly vanishes near and at the fixed point. In what follows, it
therefore suffices to compute the flow equations in the symmetric regime with
$\kappa = 0$.

By splitting the two-point correlator into its field-independent propagator part
$\Gamma^{(2)}_{k,0} \equiv \Gamma^{(2)}_{k}|_{\tilde\rho=\Psi=\bar\Psi=0}$ at
which the effective average action becomes minimal and the part including the
(not necessarily small) fluctuations around that minimum, $\Delta \Gamma_k^{(2)}
= \Gamma_k^{(2)} - \Gamma_{k,0}^{(2)}$, we can expand the logarithm and write
\begin{align} \label{eq:wetterich_eq_expansion}
& \partial_t \Gamma_k = \frac{1}{2} \tilde \partial_t \STr \ln
\left(\Gamma_{k,0}^{(2)} + \mathrm R_k \right)
 \nonumber \\ & \quad
+ \frac{1}{2} \tilde \partial_t \STr \sum_{n=1}^{\infty} \frac{(-1)^{n+1}}{n}
\left[\left(\Gamma_{k,0}^{(2)}+\mathrm R_k \right)^{-1} \Delta \Gamma_k^{(2)}
\right]^n.
\end{align}
Plugging our ansatz Eq.~\eqref{eq:truncation} into
Eq.~\eqref{eq:wetterich_eq_expansion} and evaluating for non-vanishing but
constant fields $\Phi(p)=\Phi \delta(p)$, $\Psi(p)=\Psi \delta(p)$, we get the
beta function for the Yukawa-type coupling by comparing coefficients of the
$\phi_a \bar\Psi (\sigma_a \otimes \mathbbm{1}_4) \Psi$ terms:
\begin{align}
 \partial_t g^2 & = (D-4+\eta_\phi + 2\eta_\Psi) g^2
\nonumber \\ &\quad \label{eq:dtg2}
- 8 (S-1) v_D \ell_{11}^{\mathrm{(FB)},D}(u'(0);\eta_\Psi,\eta_\phi)\,g^4.
\end{align}
The definition of the regulator-dependent threshold function
$\ell_{11}^{\mathrm{(FB)},D}$ is again found in the Appendix. $S=0$ in the
chiral Ising model, whereas $S=2$ in the chiral Heisenberg model.

\subsection{Anomalous dimensions}

 For computing the boson (fermion) anomalous
dimensions $\eta_\phi$ ($\eta_\Psi$) we again make use of the expansion
Eq.~\eqref{eq:wetterich_eq_expansion}, which we now evaluate for non-constant
boson (fermion) field $\phi=\phi(p)$ [$\Psi=\Psi(p)$, $\bar\Psi = \bar\Psi(p)$]
and vanishing fermion (boson) field $\bar\Psi = \Psi = 0$ ($\phi = 0$), and
further expand in the momentum up to order $\mathcal O(p^2)$ [$\mathcal O(p)$].
The coefficient in front of the $p^2\phi(-p)\phi(p)$ [$\bar\Psi(p) i\gamma_\mu
p_\mu \Psi(p)$] term determines $\partial_t Z_{\phi,k}$ ($\partial_t
Z_{\Psi,k}$). With the definitions in Eqs.~\eqref{eq:anomalous_dimensions} we
obtain
\begin{align}
 \label{eq:eta-phi}
\eta_\phi & = \frac{8 d_\gamma \Nf v_D}{D} m_4^{\mathrm{(F)},D}(\eta_\Psi)\,g^2
 \shortintertext{and}
 \label{eq:eta-Psi}
\eta_\Psi & = \frac{8 (S+1) v_D}{D}
m_{12}^{\mathrm{(FB)},D}(u'(0);\eta_\Psi,\eta_\phi)\,g^2,
\end{align}
where we have employed the threshold functions $m_{12}^{\mathrm{(FB)},D}$ and
$m_4^{\mathrm{(F)},D}$ (see Appendix).

\section{Fixed points and critical exponents} \label{sec:fixed_points}
\subsection{\texorpdfstring{$(4-\epsilon)$-expansion}{(4-eps)-expansion}}
\label{sec:fixed-points_4-eps}
In $D=4-\epsilon$ space-time dimensions for small $\epsilon \ll 1$ the flow
equations simplify considerably. This provides for a non-trivial and very useful
cross-check of our computation in the previous section, when comparing with the
known flow equations obtained from standard Wilsonian or minimal substraction RG schemes.
We expand the effective potential $u(\tilde\rho)$ around its minimum at $\tilde
\rho = 0$
analogously to Eq.~\eqref{eq:expansion_effective-potential} and use the dimensionless
renormalized couplings $m^2 = u'(0) = Z_{\phi,k}^{-1} k^{-2} \bar \lambda^{(1)}_k$
and $\lambda = \frac{1}{8} u''(0) = \frac{1}{8} Z_{\phi,k}^{-2} k^{D-4} \bar \lambda^{(2)}_k$.
We have seen in Sec.~\ref{sec:functional_RG} that $D=4$ constitutes an upper critical dimension of
our theory, and we thus expect the fixed-point values of an anticipated interacting critical point
to be of the order of $m^{*2}, \lambda^*, g^{*2} = \mathcal O(\epsilon)$.
Since higher bosonic self-interactions $\sim \phi^{2n}$, $n\geq 3$, are irrelevant
near the upper critical dimension [Eq.~\eqref{eq:scaling-dimension_lambda-g}], the corresponding fixed-point couplings
would be of higher order, $u^{*(n\geq 3)}(0) = \mathcal O(\epsilon^{n-1})$.
We thus may neglect them to first order in $\epsilon$, and the same applies to
higher-derivative terms [Eqs.~\eqref{eq:scaling-dimension_higher-derivative_A}--\eqref{eq:scaling-dimension_higher-derivative_C}].

For calculational simplicity, it is convenient to employ the sharp-cutoff
regulator $\mathrm R_k^\text{sc}$, yielding the threshold functions
$\ell_0^{\mathrm{(B/F)}}(\omega;\eta_{\phi/\Psi}) = -\ln(1+\omega) +
\text{const.}$, $m_{12}^{\mathrm{(FB)},D}(\omega;\eta_\Psi,\eta_\phi) =
(1+\omega)^{-2}$, and $m_4^\mathrm{(F)}(\eta_\Psi) = 1$. A formal definition of
$\mathrm R_k^\text{sc}$ is given in the Appendix.
We have checked numerically that our predictions for the universal quantities
(such as critical exponents) are regulator-independent for $D \rightarrow 4^-$.
The reason is that our ansatz for $\Gamma_k$ [Eq.~\eqref{eq:truncation}]
is  exact to first order in $\epsilon$.
Using the rescaled couplings
\begin{align} \label{eq:coupling_rescaling}
 \lambda/(8\pi^2) & \mapsto \lambda & \text{and} && g / (8\pi^2) & \mapsto g
\end{align}
the $\beta$-functions become for $\Nf=2$
\begin{align} \label{eq:beta_functions_4-eps}
\partial_t m^2 & = (-2+\eta_\phi) m^2 - 4(S+3)\frac{\lambda}{1+m^2} + 8 g^2 , \\
\partial_t \lambda & = (-\epsilon + 2\eta_\phi)\lambda + 4(S+9)
\frac{\lambda^2}{\left(1+m^2\right)^2} - 2 g^4, \\
\partial_t g^2 & = (-\epsilon + \eta_\phi + 2\eta_\Psi ) g^2 - 2(S-1)
\frac{g^4}{1+m^2},
\end{align}
with the anomalous dimensions
\begin{align} \label{eq:eta_functions_4-eps}
 \eta_\phi & = 4 g^2,  &
 \eta_\Psi & = \frac{S+1}{2} \frac{g^2}{1+m^2}.
\end{align}
Here, we have used $d_\gamma = 4$ and $v_D = 1/(32\pi^2) + \mathcal
O(\epsilon)$. We note that
Eqs.~\eqref{eq:beta_functions_4-eps}--\eqref{eq:eta_functions_4-eps} are exactly
the one-loop results as have been found earlier within the standard Wilsonian RG
approach.\cite{Herbut:2009vu} In the sharp-cutoff scheme this is in fact even
true right up to the exact same coupling rescaling
[Eqs.~\eqref{eq:dimensionless_couplings} and \eqref{eq:coupling_rescaling}].

For completeness, let us quote the fixed-point values together with the
corresponding universal exponents which determine the critical behavior. Aside from
the fully IR repulsive Gaussian fixed point (and an assumingly unphysical zero
of the $\beta$-functions at $\lambda^*<0$) we recover the well-known
Wilson-Fisher fixed point at $g^{*2} = 0$ and $\lambda^* > 0$, being repulsive
in the $g^2$-direction, and a fermionic critical point at
\begin{align}
m^{*2} & = \frac{24}{(9+S)(7-S)}\epsilon + \mathcal O(\epsilon^2), \\
\lambda^* & = \frac{2}{(9+S)(7-S)}\epsilon + \mathcal O(\epsilon^2), \\
g^{*2} & = \frac{1}{7-S}\epsilon + \mathcal O(\epsilon^2),
\end{align}
which is attractive both in the $\lambda$- and the $g^2$-directions.
We note that the fixed-point values are of order $\mathcal
O(\epsilon)$, as anticipated. To the present order in $\epsilon$,
this confirms \emph{a posteriori} the validity
of neglecting all higher-order interactions.
Close to the
critical point the correlation length scales as $\xi \propto |\delta|^{-\nu}$,
with the ``reduced temperature'' $\delta \coloneqq m^{2} - m^{*2}$, measuring
the distance from criticality. We find
\begin{align}
 1/\nu = 2 - \frac{12(5+S)}{(9+S)(7-S)}\epsilon + \mathcal O(\epsilon^2).
\end{align}
For the anomalous dimensions at the critical point we get
\begin{align} \label{eq:eta_FP_4-eps}
 \eta_\phi & = \frac{4}{7-S}\epsilon + \mathcal O(\epsilon^2) &
 \text{and} &&
 \eta_\Psi & = \frac{1+S}{2(7-S)}\epsilon + \mathcal O(\epsilon^2),
\end{align}
with $\eta_\phi$ as the usual order-parameter's anomalous dimension, determining
the scaling of the order-parameter correlation function at the critical point,
$\langle \phi(x) \phi(y) \rangle_\text{conn.} \propto 1/|x-y|^{D-2+\eta_\phi}$,
and the fermionic anomalous dimension $\eta_\Psi$, determining $\langle \Psi(x)
\bar\Psi(y) \rangle_\text{conn.} \propto 1/|x-y|^{D-1+\eta_\Psi}$ at the
critical point, respectively. \footnote{Note that the factor $3$ in the numerator of the Eq. (13)
in Ref.\ \onlinecite{Herbut:2009vu} should be $1+S$.}

\subsection{\texorpdfstring{Numerical evaluation for $2<D<4$}{Numerical
evaluation for 2<D<4}}
For general $D \in (2,4)$ we evaluate the flow equations [Eqs.~\eqref{eq:dtu},
\eqref{eq:dtg2}--\eqref{eq:eta-Psi}] numerically. A necessary condition for
reliability of our results is that the regulator-dependences of our universal
predictions remain small when $D$ is no longer close to the upper critical
dimension. We check this requirement by employing both the sharp-cutoff scheme
as well as the linear regulator $\mathrm R_k^\text{lin}$, which is also defined
in the Appendix.
$\mathrm{R}_k^\text{lin}$ shares with the sharp regulator $\mathrm
R_k^{\text{sc}}$ the convenient property that all occurring loop integrals can
be carried out analytically. For both regulators the results for these integrals
are given in the Appendix.

\begin{figure*}
\includegraphics[width=.48\textwidth]{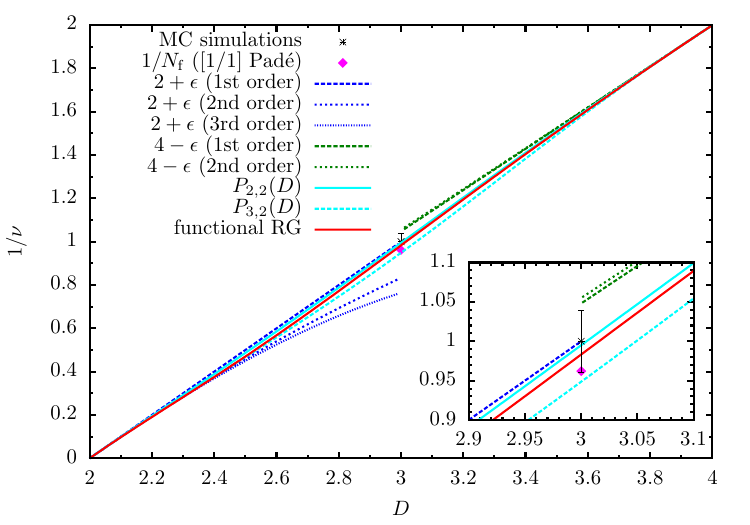}\hfill
\includegraphics[width=.48\textwidth]{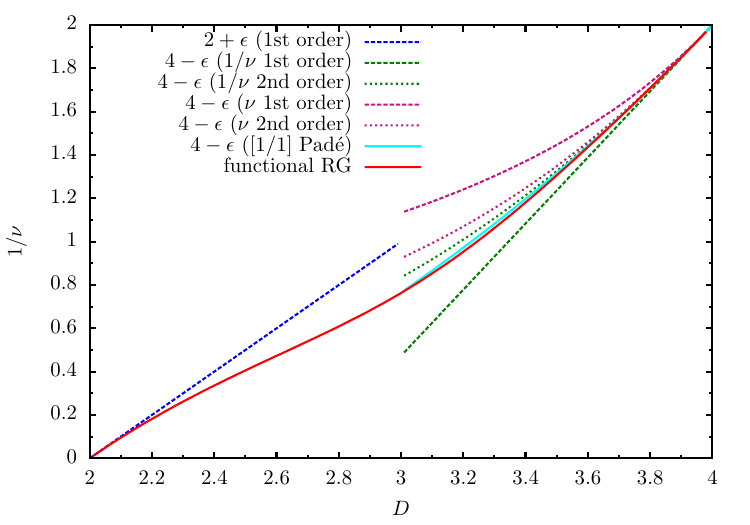}
\caption{Correlation-length exponent $1/\nu$ for chiral Ising (left panel) and
chiral Heisenberg (right panel) universality classes from functional RG with sharp
regulator (red/solid line) and for comparison from MC simulations,\cite{Karkkainen:1993ef}
2nd-order $1/\Nf$-expansion ([1/1] Pad\'e
resummed),\cite{Vasiliev:1992wr, Gracey:1993kc} 3rd-order
$(2+\epsilon)$-expansion,\cite{Gracey:1990sx, Vasiliev:1992wr} 2nd-order
$(4-\epsilon)$-expansion,\cite{Karkkainen:1993ef, Rosenstein:1993zf} and
polynomial interpolations $P_{i,j}(D)$ of $i$th-order $(2+\epsilon)$- and
$j$th-order $(4-\epsilon)$-expansion. In the
right panel we also demonstrate the ambiguity of the plain
$(4-\epsilon)$-expansion coming from either expanding $1/\nu$ or $\nu$ itself; cf.\ the discussion in Sec.~\ref{sec:discussion}.}
\label{fig:nu}
\end{figure*}

\begin{figure*}
\includegraphics[width=.48\textwidth]{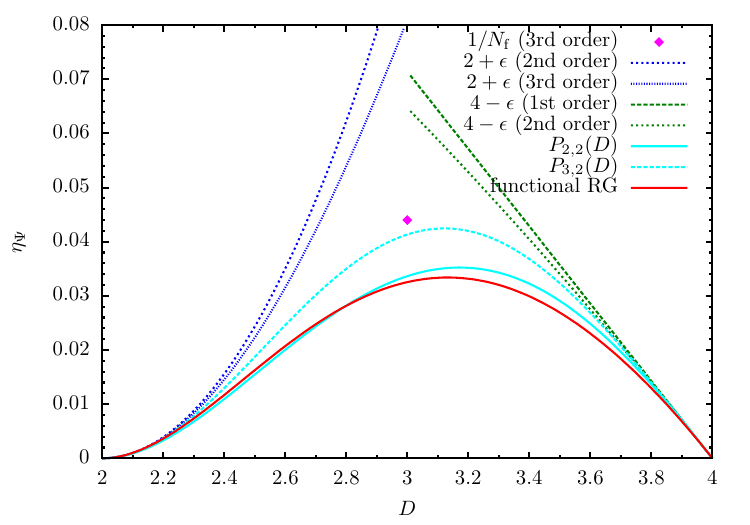}\hfill
\includegraphics[width=.48\textwidth]{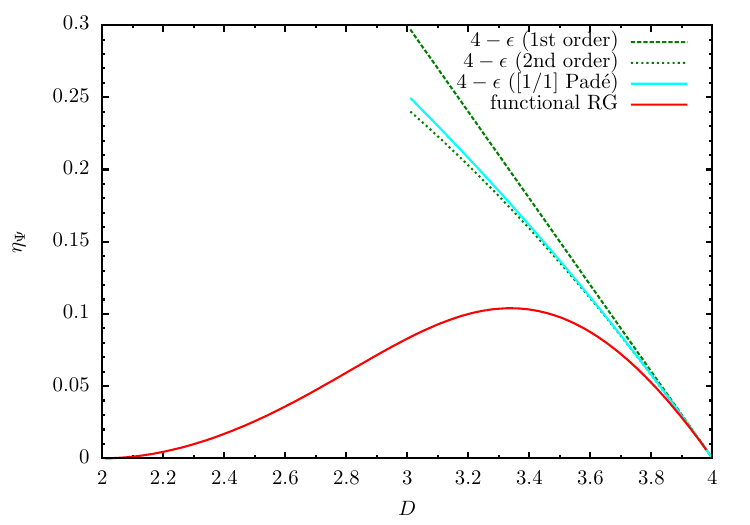}
\caption{Same as Fig.~\ref{fig:nu} for fermionic anomalous dimension
$\eta_\Psi$. Left panel: chiral Ising universality class. Right panel: chiral
Heisenberg universality class.}
 \label{fig:eta_psi}
\end{figure*}

The defining equation for the fixed-point potential $\partial_t u^*(\tilde \rho)
= 0$ is a second-order ordinary nonlinear differential equation [see
Eq.~\eqref{eq:dtu}]. For any given $g^2$, it can be solved numerically.\cite{Hofling:2002hj}
For criticality alone it is, however, just as good,
and technically much more convenient, to employ a Taylor expansion around the
potential's minimum at $\tilde\rho = 0$, as in Eq.~\eqref{eq:expansion_effective-potential}.
For our numerical results, we truncate this expansion after the 6th
order in $\tilde\rho$, i.e., we neglect all interactions $\sim \phi^{14}$ and
higher. The order of the polynomial truncation is chosen such that an inclusion
of higher-order terms changes our predictions for the critical exponents only
beyond the third digit. The error introduced by truncating the effective
potential is thus much smaller than the error we expect due to the truncation of
$\Gamma_k$, Eq.~\eqref{eq:truncation}. Our results for correlation-length
exponent $\nu$ and anomalous dimensions $\eta_\phi$ and $\eta_\Psi$ are shown in
Figs.~\ref{fig:nu}--\ref{fig:eta_psi}. For clarity, we have plotted only the
sharp-cutoff results, since the difference to the linear-regulator exponents is
hardly visible within the given resolution of these plots.
Our numerical predictions in $D=3$ are given for both regulators in
Table~\ref{tab:exponents_chiral_Ising} for the chiral Ising ($S=0$) universality
class and Table~\ref{tab:exponents_chiral_Heisenberg} for the chiral Heisenberg
($S=2$) universality class, respectively. In
Table~\ref{tab:exponents_chiral_Heisenberg}, we have also included the exponent
$\omega$, determining the leading correction to scaling, e.g., for the correlation
length $\xi \propto |\delta|^{-\nu}(1+ a_{\pm} |\delta|^{\omega\nu} + \mathcal
O(\delta^2))$. Since there does not seem to be any dangerously irrelevant
coupling in the problem, we expect hyperscaling to hold. Our predictions for the
remaining exponents $\alpha$, $\beta$, $\gamma$, and $\delta$, obtained by the
usual relations,\cite{Herbut2007modern} are given in
Table~\ref{tab:exponents_chiral_Heisenberg}, too.

\begin{figure*}
\includegraphics[width=.48\textwidth]{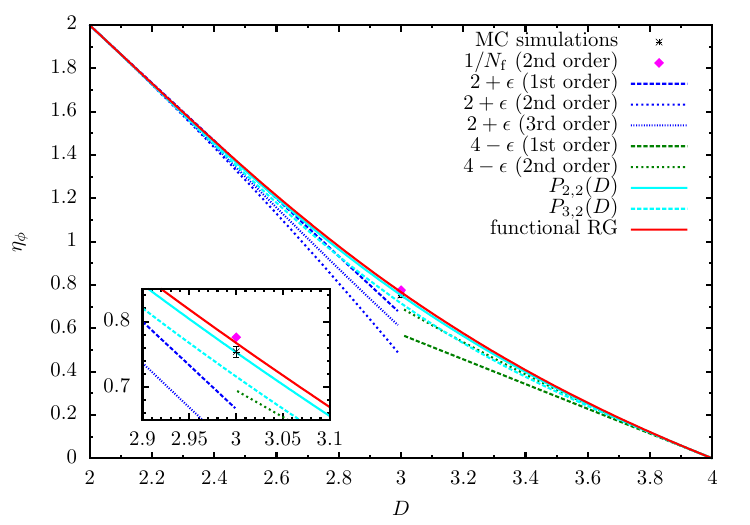}\hfill
\includegraphics[width=.48\textwidth]{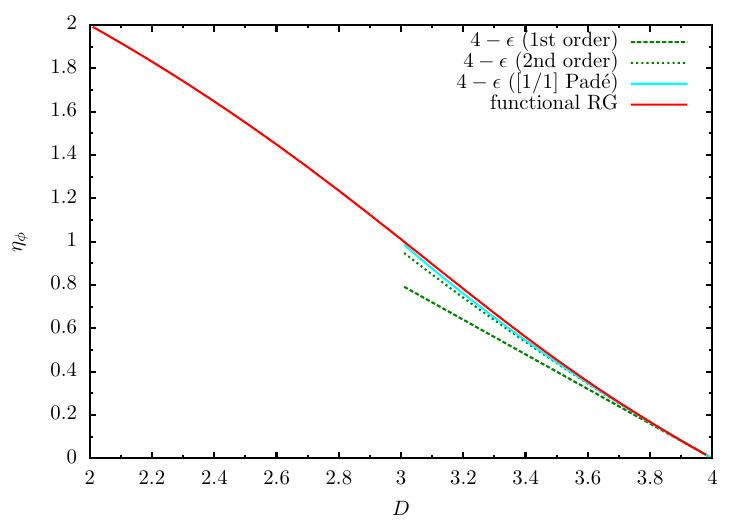}
\caption{Same as Fig.~\ref{fig:nu} for anomalous dimension of order parameter
$\eta_\phi$. Left panel: chiral Ising universality class. Right panel: chiral
Heisenberg universality class.}
\end{figure*}

\section{Discussion} \label{sec:discussion}
Due to the absence of an obvious small expansion parameter in the
strongly-coupled system for general $D\in(2,4)$, the truncation-induced error is
hard to control. However, since the chiral Ising universality class is by now
fairly well-established it provides a useful testing ground to check the
reliability of our approximation. Assuming similar performances in the two
universality classes, we can therewith estimate the accuracy of our predictions
in the chiral Heisenberg universality class.

\begin{table}
\caption{Critical exponents in $D=3$ for the transition into the
charge-density-wave state (chiral Ising universality class, $S=0$, with
$d_\gamma \Nf = 8$) from different methods. Functional RG results (this work) in
LPA' approximation and by truncating $u(\tilde\rho)$ after 6th order in
$\tilde\rho$, both for sharp ($\mathrm R_k^\text{sc}$) and linear regulator
($\mathrm R_k^\text{lin}$). Previous FRG results without truncating
$u(\tilde\rho)$. $P_{i,j}(D)$ interpolates between $i$th-order
$(2+\epsilon)$-expansion and $j$th-order $(4-\epsilon)$-expansion results, see
Sec.~\ref{sec:discussion}.}
\label{tab:exponents_chiral_Ising}
\begin{tabular*}{.48\textwidth}{@{\extracolsep{\fill}}lccc}
\hline\hline
& $1/\nu$ & $\eta_\phi$ & $\eta_\Psi$ \\
\hline
FRG [LPA', $\mathcal O(\tilde\rho^6)$, $\mathrm R_k^\text{lin}$] & 0.982 & 0.760
& 0.032 \\
FRG [LPA', $\mathcal O(\tilde\rho^6)$, $\mathrm R_k^\text{sc}$] & 0.978 & 0.767
& 0.033 \\
FRG [LPA', full $u(\tilde\rho)$, $\mathrm R_k^\text{lin}$]\cite{Hofling:2002hj}
& 0.982 & 0.756 & 0.032 \\
$1/\Nf$-expansion (2nd/3rd order) \cite{Vasiliev:1992wr, Gracey:1993kc} &
$\phantom{^*}$0.962$^*$ & 0.776 & 0.044\\
$(2+\epsilon)$-expansion (3rd order) \cite{Gracey:1990sx}& 0.764 & 0.602 & 0.081
\\
$(4-\epsilon)$-expansion (2nd order) \cite{Karkkainen:1993ef, Rosenstein:1993zf}
& 1.055 & 0.695 & 0.065 \\
Polynomial interpolation $P_{2,2}$ & 0.995 & 0.753 & 0.034 \\
Polynomial interpolation $P_{3,2}$ & 0.949 & 0.716 & 0.041 \\
Monte-Carlo simulations \cite{Karkkainen:1993ef}${}^\dagger$ & 1.00(4) &
0.754(8) & $-$ \\
\hline\hline
\end{tabular*}\\ \vspace*{-.75\baselineskip}
\begin{flushleft}
{\footnotesize %
${}^\ast$ [1/1] Pad\'{e} approximant, Eq.~\eqref{eq:Pade}.
\\
${}^\dagger$ cubic-lattice model with smaller
symmetry, sign problem ignored.\cite{Chandrasekharan:2013aya}}
\end{flushleft}
\end{table}

\subsection{Chiral Ising universality class}
Within the $1/\Nf$-expansion, the Gross-Neveu model was solved in any dimension
$2\leq D \leq 4$ up to two-loop order, with the fermion anomalous dimension
being known even up to three-loop order.\cite{Vasiliev:1992wr, Gracey:1993kc}
In $D=3$ the critical exponents read as
\begin{align}
1/\nu & = 1 - \tfrac{8}{3\pi^2 \Nf} + \tfrac{4(632 + 27\pi^2)}{27\pi^4 \Nf^2}
= 1 - \tfrac{0.270}{\Nf} + \tfrac{1.366}{\Nf^2}, \label{eq:theta_large-Nf} \\
\eta_\phi & = 1 - \tfrac{16}{3\pi^2\Nf} + \tfrac{4(304 - 27\pi^2)}{27\pi^4\Nf^2}
= 1 - \tfrac{0.540}{\Nf} + \tfrac{0.057}{\Nf^2}, \\
\eta_\Psi & = \tfrac{2}{3\pi^2\Nf} + \tfrac{112}{27\pi^4\Nf^2}
  + \tfrac{94\pi^2 + 216\pi^2 \ln 2 - 2268 \zeta(3) - 501}{162\pi^6\Nf^3}
 \nonumber \\
 & =  \tfrac{0.068}{\Nf} + \tfrac{0.043}{\Nf^2} - \tfrac{0.005}{\Nf^3},
\end{align}
with $\Nf$ counting the number of four-component fermion species. The exponents
have also been computed up to three-loop order within an expansion around the
lower critical dimension.\cite{Gracey:1990sx, Gracey:1991vy, Luperini:1991sv}
For $\Nf=2$ and $D=2+\epsilon$, one obtains:
\begin{align} \label{eq:chiIsing_2+eps_theta}
 1/\nu & = \epsilon - \tfrac{1}{6}\epsilon^2 - \tfrac{5}{72}\epsilon^3
         = \epsilon - 0.167 \epsilon^2 - 0.069 \epsilon^3, \\
\eta_\phi & = 2 - \tfrac{4}{3} \epsilon - \tfrac{7}{36} \epsilon^2 +
\tfrac{7}{54} \epsilon^3 \nonumber \\ &
= 2 - 1.333 \epsilon - 0.194 \epsilon^2 + 0.130 \epsilon^3, \\
\label{eq:chiIsing_2+eps_eta-psi}
 \eta_\Psi & = \tfrac{7}{72}\epsilon^2 - \tfrac{7}{432}\epsilon^3
             = 0.097 \epsilon^2 - 0.016 \epsilon^3.
\end{align}
Here, the anomalous dimensions are known even to four-loop order.
\cite{Vasiliev:1997sk, Gracey:2008mf}

The corresponding partially bosonized system, the Gross-Neveu-Yukawa model, was
solved to two-loop order in $D=4-\epsilon$ dimensions with (for $\Nf=2$)
\cite{Karkkainen:1993ef, Rosenstein:1993zf}\textsuperscript{,}\footnote{Let us
point out some typing errors in the formulas for $\gamma_\psi$ and
$\bar\gamma_{\phi^2}$ as given in Eqs.~(11) and (12) of the work by Rosenstein
\emph{et al.}:\cite{Rosenstein:1993zf} In Eq.~(12) in the second term of the two-loop
coefficient it must read as $-94 N^2$ instead of $+94 N^2$ which can be seen by
comparing with the formulas given by Karkkainen \emph{et al.};\cite{Karkkainen:1993ef}
additionally, only $-94 N^2$ gives the correct $1/N$ expansion as quoted
following Eq.~(12) in Ref.~\onlinecite{Rosenstein:1993zf}. In Eq.~(11) of
Ref.~\onlinecite{Rosenstein:1993zf} the one-loop coefficient misses a factor of
$1/2$, as can be seen by comparing again with the expansion in $1/N$, following
Eq.~(11), or with Eq.~\eqref{eq:eta_FP_4-eps} in the present work. Also,
the two-loop coefficient as given in Eq.~(11) cannot be correct, since it does
not produce the correct $1/N$ expansion; we expect that in the numerator it
should read $3N$ instead of $33N$, which does the job. The quoted results in the
present work use these corrections.}
\begin{align}
 1/\nu & = 2 - \tfrac{20}{21}\epsilon + \tfrac{325}{44982}\epsilon^2
         = 2 - 0.952 \epsilon + 0.007 \epsilon^2, \label{eq:theta_up-dim}\\
 \eta_\phi & = \tfrac{4}{7}\epsilon + \tfrac{109}{882} \epsilon^2
             = 0.571 \epsilon + 0.124 \epsilon^2, \label{eq:eta-phi_up-dim} \\
 \eta_\Psi & = \tfrac{1}{14}\epsilon - \tfrac{71}{10584} \epsilon^2
             = 0.071 \epsilon - 0.007 \epsilon^2.
\end{align}
The relationship between the Gross-Neveu model in $D=2+\epsilon$ and the
Gross-Neveu-Yukawa model in $D=4-\epsilon$ is similar to the one between the
nonlinear sigma model and the Ginzburg-Landau-Wilson theory (linear sigma
model):\cite{Herbut2007modern} universality suggests that the two systems in
fact describe the same critical point, just from different sides of the
transition. Indeed, when further expanding the $(4-\epsilon)$-Gross-Neveu-Yukawa
exponents in $1/\Nf$, one finds that the coefficients are order by order the
same as those one would get by expanding the $1/\Nf$-Gross-Neveu exponents at
$D=4-\epsilon$. We also note that the same is true for the
$(2+\epsilon)$-expansion exponents, as expected.

The chiral Ising universality class has also been investigated within previous
FRG calculations,\cite{PhysRevLett.86.958, Hofling:2002hj, PhysRevD.83.085012}
which in some cases \cite{Hofling:2002hj} do not rely on an expansion of the
effective potential $u(\tilde\rho)$ as in
Eq.~\eqref{eq:expansion_effective-potential}, but solve the full equation for
$u(\tilde\rho)$, Eq.~\eqref{eq:dtu}. There also exist Monte-Carlo simulations on
a cubic lattice employing the staggered-fermion formulation.\cite{Karkkainen:1993ef}
Recent analyses \cite{Chandrasekharan:2013aya}, however,
suggest that these should be taken with caution for the following reasons:
Firstly, the microscopic symmetry of the cubic-lattice theory in the simulations
with single species of staggered fermions (which due to fermion doubling
corresponds to $\Nf=2$ four-component continuum fermions) is (besides phase
rotations) $\mathrm{SU}(2)\times \mathbbm{Z}_2$ and it is not clear, whether the
continuum symmetry $\mathrm{SU}(2)_\text{sp} \times \mathrm{SU}(2)_\chi \times
\mathbbm{Z}_2$ is restored close to the critical point. It is therefore not excluded
that the cubic-lattice model describes a different universality class.
Secondly, the standard auxiliary field approach in the staggered-fermion
formulation on the cubic lattice suffers from a sign problem for $\Nf=2$, which
was ignored in Ref.~\onlinecite{Karkkainen:1993ef}. A recently suggested
approach that solves the sign problem also gives \emph{different} critical exponents
in a similar model with the same (smaller-than-continuum) symmetry group of
the cubic-lattice model considered in Ref.~\onlinecite{Karkkainen:1993ef}.\cite{Chandrasekharan:2013aya}
The apparent consistence of the quoted
cubic-lattice MC measurements with the continuum predictions (see
Table~\ref{tab:exponents_chiral_Ising}) might therefore be purely coincidental.
This deserves further investigation.

$\epsilon$- or $1/\Nf$-expansions are at best asymptotic series, and a simple
extrapolation to the physical case $\epsilon=1$ and $\Nf=2$ is quite
problematic. This is particularly evident in the
correlation-length exponent in the $1/\Nf$-expansion, where for $\Nf=2$ we get
$1/\nu = 1 - 0.135 + 0.341 + \mathcal O(1/\Nf^3)$, i.e., the second-order
correction is in fact larger than the first-order correction, with no sign of
convergence. To the present order the $(2+\epsilon)$-
and the $(4-\epsilon)$-expansions are still decreasing. However, at least in
parts the (superficial) convergence is rather slow, in particular for the
anomalous dimensions. This is in contrast to the universality class' purely
bosonic equivalent, the Ising class, where the second-order
$(4-\epsilon)$-expansion gives exponents which agree with the best known values
within an error range of less than 1\%.\cite{Herbut2007modern}
Moreover, in Eqs.~\eqref{eq:theta_large-Nf} and \eqref{eq:theta_up-dim} we have
given the correlation-length exponent in each case in terms of an expansion of
$1/\nu$ instead if $\nu$ itself, which is convenient in order to compare with the
$(2+\epsilon)$-expansion, in which $\nu \propto 1/\epsilon$. A naive extrapolation
to the physical case, however, leaves us with the ambiguity of either directly
extrapolating the expansion of $1/\nu$ or first expanding $\nu$ itself and
extrapolating $\epsilon \to 1$ afterwards. Because of the comparatively large loop corrections
the difference between these two, probably equally justified, procedures
is not negligible, e.g., of the order of $10\%$ for the second-order
$(4-\epsilon)$-expansion. All this indicates the crucial necessity of
resummation of the expansions in the present fermionic systems.
Standard Borel-type of resummation techniques rely on the knowledge of the
large-order behavior of the coefficients, obtained within, e.g., a semiclassical
analysis.\cite{kleinert2001critical} As far as we are aware, no such knowledge exists yet
in the fermionic systems considered here.

\begin{figure*}
\includegraphics[width=0.33\textwidth]{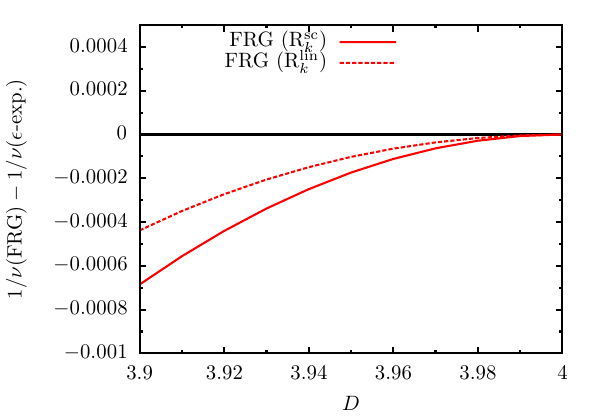}\hfill
\includegraphics[width=0.33\textwidth]{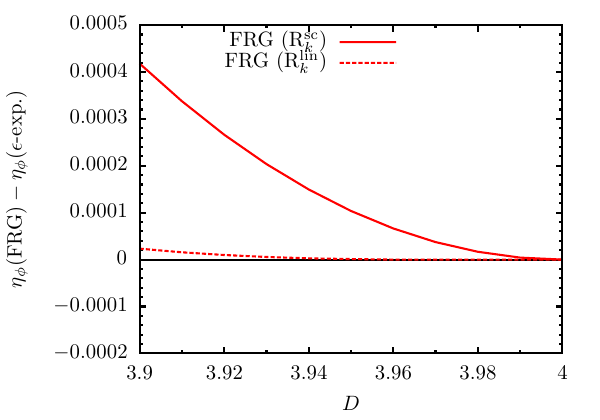}\hfill
\includegraphics[width=0.33\textwidth]{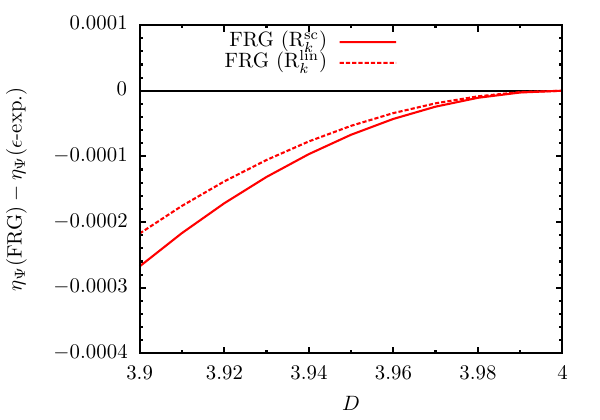}
\caption{Absolute difference of critical exponents from FRG with sharp and
linear cutoff, respectively, to $\epsilon$-expansion results near upper critical
dimension in the chiral Ising universality class. Both sharp-cutoff and
linear-cutoff scheme become numerically exact to first order in $\epsilon$.}
\label{fig:comparison_upperD}
\end{figure*}

\begin{figure*}
\includegraphics[width=0.33\textwidth]{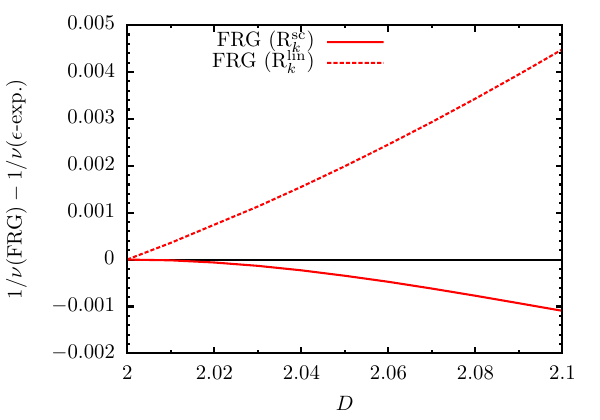}\hfill
\includegraphics[width=0.33\textwidth]{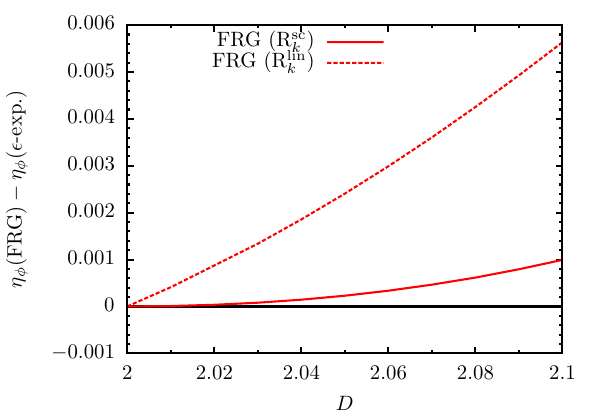}\hfill
\includegraphics[width=0.33\textwidth]{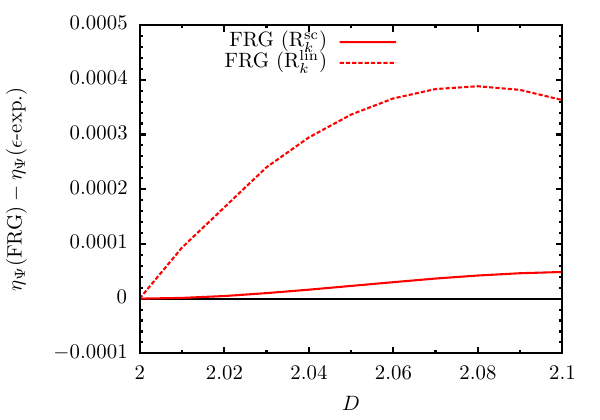}
\caption{Same as Fig.~\ref{fig:comparison_upperD} near lower critical dimension.
Here, only the sharp-cutoff scheme is numerically exact to first order
in~$\epsilon$, while the linear-cutoff scheme is exact merely to zeroth order
and yields slightly different first-order corrections.}
\label{fig:comparison_lowerD}
\end{figure*}

For the correlation-length exponent in the $1/\Nf$ expansion we therefore use a
naive symmetric $[1/1]$ Pad\'{e} approximant \cite{kleinert2001critical}
\begin{align} \label{eq:Pade}
[1/1]^{(1/\nu)}(\Nf) = \frac{584 + 27\pi^2 +18\pi^2\Nf}{632 + 27 \pi^2 + 18
\pi^2 \Nf},
\end{align}
where the coefficients have been chosen such that by again expanding in $1/\Nf$
the Pad\'e approximant gives back the original series in
Eq.~\eqref{eq:theta_large-Nf}. Although a solid justification of the simple Pad\'e
approximation is certainly out of reach, it at least solves the ambiguity between
the expansions of $1/\nu$ and $\nu$.
For the $\epsilon$-expansions, however, one can
do better: We may take advantage of the knowledge of the results from the
expansions near lower and upper critical dimension \emph{simultaneously} and try
to find a suitable interpolation between these two limits. In the purely bosonic
$\mathrm O(N)$ models, such an interpolation algorithm, based on an optimized
interpolation function within a variational approach, has been demonstrated to
yield persuasively accurate values for the critical exponents
\cite{kleinert2001critical}---even though the expansion around the lower
critical dimension yields entirely useless values when naively extrapolating to
$\epsilon=1$. In contrast, in the fermionic systems considered here, the
$(2+\epsilon)$- and the $(4-\epsilon)$-expansion yield loop corrections of
comparable order, e.g., $1/\nu \simeq \mathcal O(1)$ in $D=3$ while $1/\nu
\rightarrow 0$ ($1/\nu \rightarrow 2$) at the lower (upper) critical
dimension---a fact which makes an interpolation even more promising. To our
knowledge, such a variational resummation has so far not been pursued in the
case of the fermionic models. This deserves a study on its own. Here, instead of
employing the full optimization process, we use a simplified approach with
non-optimized interpolation function. For convenience, we employ a polynomial
interpolation $P_{i,j}(D)$ between the results from the $i$th-order
$(2+\epsilon)$-expansion and the $j$th-order $(4-\epsilon)$-expansion. This is
done by extending the $(2+\epsilon)$-expansion by $j+1$ more terms, e.g., for
the correlation-length exponent ($i=3$, $j=2$)
\begin{align}
P_{3,2}^{(1/\nu)}(D) & = (D-2) - \tfrac{1}{6} (D-2)^2 - \tfrac{5}{72} (D-2)^3 +
\nonumber \\
                     &\quad + a_4 (D-2)^4 + a_5 (D-2)^5 + a_6 (D-2)^6,
\end{align}
and fitting this extended series to the known result near the upper critical
dimension, Eq.~\eqref{eq:theta_up-dim}. I.e., we uniquely determine the
coefficients $a_{4}$, $a_5$, $a_6$ by requiring
\begin{align}
 P_{3,2}^{(1/\nu)}(4) & = 2, &
 P_{3,2}^{(1/\nu)}{}'(4) & = -\tfrac{20}{21}, &
 P_{3,2}^{(1/\nu)}{}''(4) & = \tfrac{325}{22491}.
\end{align}
By construction, the interpolating polynomial $P_{i,j}(D)$ is therefore $i$-loop
($j$-loop) exact near lower (upper) critical dimension.
This interpolational resummation also solves the ambiguity between the expansions of $1/\nu$ and
$\nu$ by construction. In order to be able to
follow the development and to compare with the symmetric case $i=j$, we have
computed $P_{i,j}(D)$ for both the second- ($i=2$) and the third- ($i=3$) order
$(2+\epsilon)$-expansion, each with the second- ($j=2$) order $(4-\epsilon)$-expansion.
The results for correlation-length exponent and anomalous dimensions are shown
in Figs.~\ref{fig:nu}--\ref{fig:eta_psi} (left panels), together with the naive
extrapolations and our FRG predictions.

\begin{table*}[tbp]
\caption{Critical exponents in $D=3$ for the transition into the
antiferromagnetic state (chiral Heisenberg universality class, $S=2$, with
$d_\gamma \Nf = 8$) from functional RG in LPA' approximation for both linear and
sharp regulator. $\alpha$, $\beta$,
$\gamma$, and $\delta$ from hyperscaling relations. For comparison: plain 2nd-order
$(4-\epsilon)$-expansion results\cite{Rosenstein:1993zf} (for both the
direct expansion of $1/\nu$ and the inverse of the expansion of $\nu$, cf.\ text) and
$[1/1]$ Pad\'{e} approximant thereof.}
\label{tab:exponents_chiral_Heisenberg}
\begin{tabular*}{\textwidth}{@{\extracolsep{\fill}}lcccccccc}
\hline\hline
& $1/\nu$ & $\eta_\phi$ & $\eta_\Psi$ & $\omega$ & $\alpha$ & $\beta$ & $\gamma$
& $\delta$\\
\hline
FRG [LPA', $\mathcal O(\tilde\rho^6)$, $\mathrm R_k^\text{lin}$] & 0.772 & 1.015
& 0.084 & 0.924 & -1.89 & 1.31 & 1.28 & 1.98 \\
FRG [LPA', $\mathcal O(\tilde\rho^6)$, $\mathrm R_k^\text{sc}$] & 0.761 & 1.012
& 0.083 & 0.908 & -1.94 & 1.32 & 1.30 & 1.98 \\
$(4-\epsilon)$-expansion ($1/\nu$ 2nd order) \cite{Rosenstein:1993zf} & 0.834 & 0.959 &
0.242 & $-$ & -1.60 & 1.17 & 1.25 & 2.06 \\
$(4-\epsilon)$-expansion ($\nu$ 2nd order) \cite{Rosenstein:1993zf} & 0.923 & 0.959 &
0.242 & $-$ & -1.25 & 1.06 & 1.13 & 2.06 \\
$(4-\epsilon)$-expansion ([1/1] Pad\'{e} approx.) & 0.765 & 0.999 & 0.252 & $-$
& -1.92 & 1.31 & 1.31 & 2.00 \\
\hline\hline
\end{tabular*}
\end{table*}

In Sec.~\ref{sec:fixed-points_4-eps} it was shown that our improved local
potential approximation within the sharp-cutoff scheme produces the correct
one-loop exponents near the upper critical dimension, and we have checked
numerically that this is also the case for the linear regulator.
This is illustrated in Fig.~\ref{fig:comparison_upperD}, where we have
plotted the absolute difference of our FRG results for both the sharp and the
linear cutoff to the 2nd-order $(4-\epsilon)$-expansion results. Indeed, the
difference as well as its derivative goes to zero as $D \to 4^-$ for both
regulators. Figure~\ref{fig:comparison_lowerD} now shows that the sharp-cutoff FRG
scheme becomes one-loop exact also near the lower critical dimension: By a
linear fit to our FRG predictions in $D=2+\epsilon$, we in fact find for the
sharp-cutoff regulator
\begin{align}
 1/\nu & = 1.00 \epsilon + \mathcal O(\epsilon^2), \\
 \eta_\phi & = 2 - 1.33 \epsilon + \mathcal O(\epsilon^2), \\
 \eta_\psi & = 0.00\epsilon + \mathcal O(\epsilon^2),
\end{align}
which are on the level of our numerical accuracy exactly the one-loop results
from the $(2+\epsilon)$-expansion, cf.\
Eqs.~\eqref{eq:chiIsing_2+eps_theta}--\eqref{eq:chiIsing_2+eps_eta-psi}. We
note, however, that the linear regulator, which is often considered as an
optimal choice,\cite{Litim:2001up} does \emph{not} produce the exact
first-order corrections near the lower critical dimension utterly: $1/\nu = 1.03
\epsilon + \mathcal O(\epsilon^2)$, $\eta_\phi = 2 - 1.29 \epsilon + \mathcal
O(\epsilon^2)$, and $\eta_\psi = 0.01\epsilon + \mathcal O(\epsilon^2)$.
Although small, the discrepancy to the exact coefficients from the
$(2+\epsilon)$-expansion is numerically significant, see
Fig.~\ref{fig:comparison_lowerD}. To our knowledge, this is the first-known example
in which the sharp-cutoff regulator yields substantially better predictions
than the linear regulator. In light of these findings we believe that
the issue of optimized RG schemes in the fermion-boson models considered here
may deserve further investigation.

The numerical estimates in $D=3$ are given for all approaches in
Table~\ref{tab:exponents_chiral_Ising}. From the size of the higher-order
corrections we expect that regarding the expansions the best estimates for the
anomalous dimensions $\eta_\phi$ and $\eta_\Psi$ are obtained from the
$1/\Nf$-series, with no need for resummation. Our FRG result for $\eta_\phi$
($\eta_\Psi$) agrees with these and with the interpolation-resummed
$\epsilon$-expansion results within the mid single-digit (lower double-digit)
percent range: $\Delta\eta_\phi/\eta_\phi \simeq 3 \dots 6\%$ and
$\Delta\eta_\Psi/\eta_\Psi \simeq 20\dots 30\%$. For the correlation-length
exponent we expect either the plain two-loop $(4-\epsilon)$-expansion or the
interpolation-resummed result to yield the most accurate value. Our FRG
prediction agrees with both within $\Delta\nu/\nu \simeq 3 \dots 7\%$. Both
$\nu$ and $\eta_\phi$ from the FRG agree with the MC measurements within an even
smaller error range: $\Delta\nu/\nu \simeq \Delta\eta_\phi/\eta_\phi \simeq
2\%$. Our findings also agree very well with the previous FRG results which
solve the full equation for the effective potential, suggesting that our
polynomial truncation, Eq.~\eqref{eq:expansion_effective-potential}, should be
just as good on our level of accuracy.
We also find that our FRG predictions only slightly depend on the specific
regulator function, which is additionally reassuring.

\subsection{Chiral Heisenberg universality class}

One might expect similar performances of our approximation in the chiral Ising and the
chiral Heisenberg universality class. For the chiral Heisenberg universality class, there are much fewer
previous calculations available; there exists, however, a two-loop calculation
in $D=4-\epsilon$ dimensions, yielding the exponents\cite{Rosenstein:1993zf}
\begin{align}
 1/\nu & = 2 - \tfrac{84}{55} \epsilon + \tfrac{2286329}{6322250} \epsilon^2
	 = 2 - 1.527 \epsilon + 0.362 \epsilon^2, \\
 \eta_\phi & = \tfrac{4}{5} \epsilon + \tfrac{4819}{30250}\epsilon^2
	     = 0.8 \epsilon + 0.159 \epsilon^2, \\
 \eta_\Psi & = \tfrac{3}{10} \epsilon - \tfrac{6969}{121000}\epsilon^2
	     = 0.3 \epsilon - 0.058 \epsilon^2,
\end{align}
with again only a slow (superficial) convergence in comparison to its purely
bosonic equivalent, the Heisenberg model.
Again, the ambiguity between either expanding $1/\nu$ or inverting the expansion
of $\nu$ itself is of the order of $10 \%$, see
Table~\ref{tab:exponents_chiral_Heisenberg}. At first order, it is yet
considerably higher.
For comparison, we have therefore also
calculated $[1/1]$ Pad\'e approximants, analogous to Eq.~\eqref{eq:Pade}. They
are plotted together with the plain $\epsilon$-expansion results and our
sharp-cutoff FRG predictions in Figs.~\ref{fig:nu}--\ref{fig:eta_psi} (right
panels).
In the case of the correlation-length exponent, we compare with both the direct
expansion of $1/\nu$ as well as the inverse of the expansion of $\nu$ itself,
in order to demonstrate the ambiguity.
The numerical estimates are given in
Table~\ref{tab:exponents_chiral_Heisenberg}. Again, we find that our FRG
approximation carries only a minor regulator dependence. $\nu$ and $\eta_\phi$
agree well with the Pad\'e-resummed $\epsilon$-expansions within $\lesssim 2\%$.
$\eta_\phi$ agrees also with the plain second-order $\epsilon$-expansion
within $\simeq 5\%$, while $\nu$ agrees only within a somewhat larger error range
$\simeq 10 \dots 20\%$, depending on whether we expand $\nu$ or $1/\nu$
in $\epsilon$.
The predictions for $\eta_\Psi$ differ to about a factor of~$3$ between FRG
and $\epsilon$-expansion---in
full analogy to the chiral Ising case, where the naive extrapolation of the
$(4-\epsilon)$-expansion does not agree well with either FRG or the
interpolational-resummation results. For completeness, we have also noted in
Table~\ref{tab:exponents_chiral_Heisenberg} our estimate for the
corrections-to-scaling exponent $\omega$ and the exponents $\alpha$, $\beta$,
$\gamma$, and $\delta$, which are related to $\nu$ and $\eta_\phi$ by the
hyperscaling relations.\cite{Herbut2007modern}

In contrast to the satisfactory agreement of our FRG predictions with those
of the second-order $\epsilon$-expansion, they appear to significantly contradict
the numerical findings of the simulations of the Hubbard model on the honeycomb
lattice: In Ref.~\onlinecite{Assaad:2013xua} an excellent collapse of the
finite-size-scaling data is obtained by assuming $\beta = 0.79$ and $1/\nu = 1.13$,
which happen to be the values from the plain first-order $(4-\epsilon)$-expansion (using the
extrapolation of $\nu$ itself).\cite{Herbut:2009vu} The exponents are $\sim 50\%$
off from our FRG predictions, and it is unlikely that a nearly as good
finite-size scaling of the lattice data would be possible with our results for
$\beta$ and $\nu$. Ref.~\onlinecite{2012NatSR...2E.992S} reports $\beta \approx
0.8$, which is close to the above quoted values, and again in clear numerical
conflict with our findings. Evidently, further analytical and numerical studies would be desirable
in order to pin down the values of the exponents in this universality class.

\section{Conclusions} \label{sec:conclusions}
In conclusion, we have investigated the Mott transition on the honeycomb lattice
from the semimetallic phase into the charge-density wave state and into the
antiferromagnetic state, respectively within an effective field-theory approach.
In the Hubbard-like model, the former transition is
expected for large nearest-neighbor repulsion, while the latter is induced by a
strong on-site repulsion.\cite{PhysRevLett.97.146401, Assaad:2013xua} They
are effectively described by the chiral Ising ($=$ $\mathbbm
Z_2$-Gross-Neveu) model and the chiral Heisenberg ($=$
$\mathrm{SU}(2)$-Gross-Neveu) model. We have employed the functional
renormalization group formulated in terms of the Wetterich equation to
compute the critical exponents, describing the critical behavior near the
continuous transition. In the chiral Ising case, our predictions, made within
the LPA' truncation of the derivative expansion, agree well with existing
calculations up to the mid single-digit percent range for $\nu$ and $\eta_\phi$
and the lower double-digit percent range for $\eta_\Psi$. We would expect a similar
accuracy in the chiral Heisenberg case, where not as many previous results exist.
However, while our predictions are in agreement with the second-order
$(4-\epsilon)$-expansion results of the chiral Heisenberg model, the significant
numerical mismatch to the measurements in the Hubbard-model simulations are
worrisome. These discrepancies may point to an as yet hidden subtlety in our
effective Gross-Neveu-Yukawa approach, or in both our FRG approximation as well
as the second-order $(4-\epsilon)$-expansion. This issue needs
clarification in future studies.
Within the FRG, a systematic improvement of the present approximation would be to incorporate the effect of newly generated four-fermion terms, e.g., by dynamical bosonization, \cite{Gies:2001nw} or to go beyond LPA' by including the higher-derivative terms from Eqs.~\eqref{eq:higher-order-ops_A}--\eqref{eq:higher-order-ops_B}.

Beyond its physical (and possibly technological) importance in the context of
graphene, we believe that the universality classes presented in this work are an
ideal testing ground to investigate the validity of nonperturbative
approximation schemes, setting the stage for quantitative comparisons between
field-theoretical tools. For the chiral Ising universality, we have shown that
our FRG results are able to compete with the most accurate predictions from all
existent other approaches.
Near the upper critical dimension we have demonstrated that our predictions
become universal and exact to one-loop order. This was of course to be expected,
since the effective Gross-Neveu-Yukawa models considered here become
perturbatively renormalizable in four space-time dimensions. In two dimensions,
in contrast, these fermion-boson theories are perturbatively not directly
accessible (only the purely fermionic Gross-Neveu models are) and a loop
expansion directly in two dimensions should be expected to be highly scheme
dependent. However, here we have demonstrated that our FRG exponents in the
Gross-Neveu-Yukawa model become universal and exact also in the limit of two
dimensions. Apparently, our nonperturbative LPA' truncation ``knows'' about the existence of
the purely fermionic Gross-Neveu model with its lower critical dimension of
two---in contrast to the conventional loop expansion. Near and above two dimensions, we
find slight scheme dependencies. However, within the sharp-cutoff scheme, our
approximation is still one-loop exact. At general dimension between lower and
upper critical dimension, the FRG yields a reasonable interpolation between
these two exact limits.

\appendix
\begin{acknowledgments}
We thank Shailesh Chandrasekharan and Holger Gies for helpful explanations and
discussions and Shuai Yin for comments on the manuscript. LJ acknowledges support by the DFG under JA\,2306/1-1, GRK\,1523,
and FOR\,723. IFH is supported by the NSERC of Canada.
\end{acknowledgments}

\section*{Appendix: Regularized loop integrals}
In this Appendix we give the details of the regularized loop integrations
occurring in the derivation of our FRG flow equations. The details of the
regularization scheme are encoded in the regulator functions
$R_k^{(\mathrm{B/F})}$, which may be expressed in terms of the dimensionless
shape functions $r_k^{(\mathrm{B/F})}$ as
\begin{align}
R_k^\mathrm{(B)}(q) & =  Z_{\phi,k} q^2 r_k^\mathrm{(B)}(q^2), &
R_k^\mathrm{(F)}(q) & =  Z_{\Psi,k} i \slashed{q} r_k^\mathrm{(F)}(q^2).
\end{align}
The Wetterich equation \eqref{eq:wetterich_eq} has a one-loop structure, and the
flow equations can therefore always be written in terms of one-loop Feynman
diagrams. The occurring single integrals define the threshold functions; as used
in this work, they are given by \cite{Berges:2000ew}
\begin{align} \label{eq:threshold-fcts_1}
\ell_{0}^{(\mathrm{B/F}),D} (\omega;\eta_{\phi/\Psi}) & = \frac{1}{2}k^{-D}
\tilde\partial_t \int_0^\infty \mathrm d x\, x^{D/2 - 1}
\nonumber \\ & \quad
\times \ln \left[P^\mathrm{(B/F)}_k(x) + \omega k^2\right]\\
\ell_{1,1}^{(\mathrm{FB}),D} (\omega;\eta_\Psi,\eta_\phi) & = - \frac{1}{2}
k^{4-D} \tilde \partial_t
\int_0^\infty \mathrm d x\, x^{D/2 - 1}
\nonumber \\ & \quad
\times \left[P^\mathrm{(F)}_k(x)\right]^{-1} \left[P^\mathrm{(B)}_k(x) + \omega
k^2\right]^{-1}\\
m_4^{(\mathrm F),D} (\eta_\Psi) & = - \frac{1}{2} k^{4-D} \tilde \partial_t
\int_0^\infty \mathrm d x\, x^{D/2 + 1}
\nonumber \\ & \quad
\times \left[\partial_x \frac{1}{x\bigl(1+r_k^\mathrm{(F)}(x)\bigr)} \right]^2\\
\label{eq:threshold-fcts_4}
m_{1,2}^{(\mathrm{FB}),D}(\omega;\eta_\Psi,\eta_\phi) & = \frac{1}{2} k^{4-D}
\tilde \partial_t
\int_0^\infty \mathrm d x\, x^{D/2}
\nonumber \\ & \quad
\times \frac{1}{x\bigl(1+r_k^\mathrm{(F)}(x)\bigr)} \partial_x
\frac{1}{P^\mathrm{(B)}_k(x) + \omega k^2}
\end{align}
where we have abbreviated the momentum-dependent parts of the inverse
regularized propagator by
\begin{align}
P^\mathrm{(B)}_k (x) & \coloneqq x \bigl( 1 + r_k^\mathrm{(B)}(x) \bigr), &
P^\mathrm{(F)}_k (x) & \coloneqq x \bigl( 1 + r_k^\mathrm{(F)}(x) \bigr)^2,
\end{align}
with $x \equiv q^2$. The scale-derivative $\tilde\partial_t$ acts only on the
regulator's $t$-dependence, which implicitly occurs by means of the regularized
propagator parts $P^\mathrm{(F/B)}_k$. It is formally defined in
Eq.~\eqref{eq:tilde-partial_t} in the main text. The prefactors $\propto
k^\alpha$ in Eqs.~\eqref{eq:threshold-fcts_1}--\eqref{eq:threshold-fcts_4} are
chosen such that the threshold functions become dimensionless.

Let us consider a one-parameter family of regulator functions, which we define
in terms of their corresponding regularized inverse propagator parts
\begin{align}
P_{k,a}^\mathrm{(B)}(q^2) = P_{k,a}^\mathrm{(F)}(q^2) =
\begin{cases}
a k^2 + (1-a) q^2, & \text{for } q^2 < k^2, \\
q^2, & \text{for } q^2 \geq k^2,
\end{cases}
\end{align}
with parameter $0 < a \leq \infty$. These regulators do not affect the fast
modes with $|q| > k$ at all, these modes thus give no contribution to the
threshold integrals after taking the $\tilde\partial_t$-derivative.
Modes below but sufficiently near the RG scale $k$ are for finite $a < \infty$ only
slightly suppressed, while deep IR modes with $|q| \ll k$ are always strongly
suppressed.

There are two representatives of this family of regulators, for which the
threshold integrals can be carried out analytically: For $a=1$, the regularized
propagator becomes constant for slow modes with $|q| < k$, turning the
integrands in Eqs.~\eqref{eq:threshold-fcts_1}--\eqref{eq:threshold-fcts_4} into
simple monomials in $x$. This defines the linear regulator,\cite{Litim:2001up}
for which the threshold functions become \cite{Hofling:2002hj}
\begin{align}
\ell_{0; \text{ lin}}^{(\mathrm{B/F}),D} (\omega;\eta_{\phi/\Psi}) & =
\frac{2}{D} \left(1-\frac{\eta_{\phi/\Psi}}{D+\tfrac{3\pm 1}{2}} \right)
\frac{1}{1+\omega}, \displaybreak[0] \\
\ell_{1,1; \text{ lin}}^{(\mathrm{FB}),D} (\omega;\eta_\Psi,\eta_\phi) & =
\frac{2}{D} \left[ \left(1-\frac{\eta_{\Psi}}{D+1} \right) \frac{1}{1+\omega}
+\right. \nonumber \\
&\left. \quad \ + \left(1-\frac{\eta_\phi}{D+2} \right) \frac{1}{(1+\omega)^2}
\right], \displaybreak[0] \\
m_{4; \text{ lin}}^{(\mathrm F),D} (\eta_\Psi) & = \frac{3}{4} +
\frac{1-\eta_\Psi}{2(D-2)},
\displaybreak[0] \\
m_{1,2; \text{ lin}}^{(\mathrm{FB}),D}(\omega;\eta_\Psi,\eta_\phi) & =
\left(1-\frac{\eta_\phi}{D+1}\right) \frac{1}{(1+\omega)^2}.
\end{align}
For large $a \gg 1$, only the modes in the thin momentum shell $[k-\delta k, k]$
with $\delta k \ll k$ give significant contributions to the threshold functions,
since all lower modes are suppressed by at least $1/a$. In the sharp-cutoff
limit $a \to \infty$, understood to be taken \emph{after} the integration over
the loop momentum $x$ and the $\tilde\partial_t$-derivative in
Eqs.~\eqref{eq:threshold-fcts_1}--\eqref{eq:threshold-fcts_4}, the threshold
functions then become \cite{PhysRevD.86.105007}
\begin{align}
\ell_{0; \text{ sc}}^{(\mathrm{B/F}),D} (\omega;\eta_{\phi/\Psi}) & =
- \ln (1+\omega) + \ell_0^\mathrm{(B/F),D}(0;\eta_{\phi/\Psi}), \\
\ell_{1,1; \text{ sc}}^{(\mathrm{FB}),D} (\omega;\eta_\Psi,\eta_\phi) & =
\frac{1}{1+\omega}, \displaybreak[0] \\
m_{4; \text{ sc}}^{(\mathrm F),D} (\eta_\Psi) & = 1, \\
m_{1,2; \text{ sc}}^{(\mathrm{FB}),D}(\omega;\eta_\Psi,\eta_\phi) & =
\frac{1}{(1+\omega)^2}.
\end{align}
%


\end{document}